\documentclass[review]{elsarticle}

\usepackage{bm,amssymb,amsmath,mathrsfs}
\usepackage{amsthm}
\usepackage[usenames]{color}
\usepackage[colorlinks]{hyperref}

\usepackage{dcolumn}
\usepackage{graphicx}

\begin{document}

\begin{frontmatter}

\title{Radiative Spin Polarization in High Energy Storage Rings}

\author[cc]{S.~R.~Mane}
\ead{srmane001@gmail.com}
\address[cc]{Convergent Computing Inc., P.~O.~Box 561, Shoreham, NY 11786, USA}
  
\begin{abstract}
The usual theoretical model for synchrotron radiation in circular accelerators (synchrotrons and storage rings)
is to treat a single electron moving in a horizontal circle in a uniform vertical magnetic field,
but the true situation in real storage rings is more complicated and exhibits much richer physics.
The magnetic fields are inhomogeneous, and there is a bunch of many particles 
and they traverse a distribution of orbits (hence they encounter different magnetic fields).
This results in so-called ``depolarizing spin resonances'' (which do not appear in a simple model of a uniform vertical magnetic field). 
The calculation of the equilibrium electron spin polarization requires a much more careful analysis.
For example, a key insight is that, for motion in inhomogeneous magnetic fields, ``spin flip'' is in general \emph{not} a $180^\circ$ reversal of the spin orientation.
The physics of radiative spin polarization involves a mix of many disciplines, and provides a good example of cross-disciplinary thinking.
We shall also briefly note the connection to astrophysics.
The astrophysics literature mainly treats electron motion in very strong magnetic fields, stronger than the Schwinger critical field
(for example a neutron star).
It is a problem of ongoing interest in astrophysics to study the radiation by electrons circulating in such strong magnetic fields.
This article aims to provide the reader with a survey of the basic physics principles of radiative spin polarization,
omitting low-level mathematical algebra as much as possible.
Such details can be found in the literature, and are not relevant here.
\end{abstract}

\vskip 0.25in
\end{frontmatter}

\tableofcontents

\begin{center}
\emph{Dedicated to the memory of Robert G.~Glasser, teacher and friend.}
\end{center}

\section{Introduction}\label{ch:intro}
When electrons or positrons circulate in high-energy storage rings,
their spins become spontaneously polarized by the emission of synchrotron radiation.
This phenomenon is called radiative spin polarization.
Although synchrotron radiation is usually visualized in terms of classical physics, it \emph{does} consist of the emission of photons.
Radiative spin polarization occurs because the electron/positron spin operator couples to the photon radiation field.

Measuring the spin precession frequency is the most accurate known technique to calibrate the beam energy of a storage ring.
This requires the spins of the circulating particles to be polarized.
The major application of radiative spin polarization to date has been accurate beam energy calibration.
This technique has been employed to accurately measure the masses of numerous vector mesons produced in $e^+e^-$ annihilation experiments.
See the review by \cite{MSY2} for details.
At CERN, the mass of the $Z^0$ boson ($91.1876$ GeV) was measured to an accuracy of $\pm2.1$ MeV \cite{PDG_Z0}.
The measurement of the mass of the $Z^0$ boson was one of the highlights of the LEP physics program \cite{Zedometry}.
See also \cite{LEPenergy} for a review of the precise calibration of the LEP beam energy.

HERA operated as a high-energy lepton-proton collider in Hamburg, Germany.
So-called ``spin rotators'' \cite{BuonSteffen} were installed in the HERA lepton ring,
to rotate the polarization from vertical in the ring arcs to longitudinal at the interaction points.
See \cite{BarberHERApol} for a report on the high degree of radiative spin polarization attained in the HERA lepton ring.
HERA remains the only storage ring to deliver longitudinally polarized positrons for HEP experiments.
See, for example, an overview of results from the HERMES experiment \cite{HERMESoverview}.
\cite{Vasserman_etal_1987} employed measurements using radiative spin polarization to compare the anomalous magnetic moments
of counterpropagating electron and positron beams in the VEPP-2M storage ring at the Budker Institute of Nuclear Physics in Novosibirsk,
obtaining equality to a precision of $1\times10^{-8}$.
It was at the time the most precise verification of CPT invariance for leptons.

The theory of synchrotron radiation has its origins in the developments of classical electrodynamics in the late $19^{th}$ and early $20^{th}$ centuries,
when it was found that an accelerated charged particle emits radiation. 
If the orbit of the particle is a circle, the emission is called synchrotron radiation.
The culmination of the theory was the work of \cite{Schott}.
However, the quantum theory was being born: the solar system model of the atom had been proposed by Rutherford
but electrons in atoms do not emit synchrotron radiation.
Indeed the non-observation of synchrotron radiation in atoms was one of the problems of classical physics which gave rise to the quantum theory.
It was not until after World War 2 that electron synchrotrons of sufficiently high energy were built and synchrotron radiation was observed.
(Schott died in 1937.)

Bohr's correspondence principle applies for electron orbits of large radius and high energy,
hence a classical approximation is valid for ultrarelativistic electrons in synchrotrons and storage rings.
The modern treatment of classical synchrotron radiation follows \cite{Schwinger1949}.
However, synchrotron radiation \emph{does} consist of the emission of discrete photons,
and the classical formula is only the first term in a quantum-theoretic perturbation series.
\cite{Schwinger1954} calculated the leading-order $O(\hbar)$ corrections to the synchrotron radiation power spectrum, for unpolarized electrons.

Subsequent authors extended the theory to higher orders in $\hbar$ and also took into account the spin orientation of the electrons.
As with Schott and classical synchrotron radiation,
radiative spin polarization is a phenomenon which was predicted theoretically before it was observed experimentally.
\cite{TLK} calculated that the difference of the spin-flip radiation intensities (for spin-flips up-down and down-up) was nonzero,
which implied that an initially unpolarized circulating electron beam would become spontaneously polarized by the emission of spin-flip photons.
\cite{ST} performed the complete calculation of the radiated power spectrum, to $O(\hbar^2)$,
and also the spin-flip probabilities per unit time.
They treated a model of ultrarelativistic motion in a horizontal circle of fixed radius $\rho$ in a uniform vertical magnetic field, using the Dirac equation.
\cite{ST} found that the equilibrium polarization level in their idealized model is very high, viz.~$8/(5\sqrt3)\;(\simeq 92.4\%)$.
The self-polarization of stored electron and positron beams is now called the ``Solokov-Ternov'' effect, in their honor.

Radiative spin polarization is a slow process.
The spin is of $O(\hbar)$ and the spin-flip probabilities per unit time are of $O(\hbar^2)$.
The spin-flip radiated power is too small to observe with present-day detectors,
but it leaves its imprint on the electron spins, and the polarization of the electron spins can be measured.
The first experimental observation of radiative spin polarization was at Orsay,
at the storage ring ACO \cite{Orsay1971} and at the Budker Institute for Nuclear Physics, at the storage ring VEPP-2 \cite{Baier1972}.
Later, a higher degree of polarization was obtained at ACO \cite{LeDuff_etal} and at VEPP-2M \cite{VEPP2M_radpol}.

This article is mainly concerned with electrons circulating in storage rings
but the problem is also of interest in astrophysics (where it is called ``cyclotron radiation'').
The astrophysics literature mainly treats electron motion in very strong magnetic fields, stronger than the Schwinger critical field (for example a neutron star).
The Schwinger critical electric field $E_{\rm Sch}$ is defined as the electric field which accelerates an electron (in linear motion)
through its rest energy over a distance of one Compton wavelength,
i.e.~$eE_{\rm Sch}\times \hbar/(m_ec) = m_ec^2$, hence $E_{\rm Sch} = m_e^2c^3/(\hbar e)\simeq 1.32\times10^{18}\;\textrm{V/m}$.
Here $m_e$ is the electron mass and $e$ is the electron charge.
The Schwinger critical magnetic field $B_{\rm Sch}$ is defined via
$B_{\rm Sch} = E_{\rm Sch}/c = m_e^2c^2/(\hbar e) \simeq 4.41\times10^9\;\textrm{T}$.
In an early paper on the subject, \cite{Latal}
studied photon emissions from low-lying energy levels, but took into account the spin orientation of the electrons.
More recent treatments, which treat arbitrarily high-lying energy levels, are given by \cite{BaringAstro} and \cite{SemionovaAstro}.
The subject continues to be of interest in astrophysics.

\cite{JacksonRMP} published what has become a classic review article on radiative spin polarization.
Jackson treated the quantum electrodynamics of spin-flip photon emissions and electron motion in a fixed horizontal circular orbit
(or circular arcs joined by straight lines).
Here we shall expand on the subject to treat the additional physics required to deal with real high-energy electron storage rings.
This article is partly a homage to Jackson.

In particular, note the following.
The electrons in a storage ring, in particular the idealized model of circular motion in a uniform vertical magnetic field,
are essentially at zero temperature.
Hence why is the asymptotic polarization lower than $100\%$?
Quoting from \cite{JacksonRMP}:
\begin{quote}
  The reader may, with justification, feel that the author has wandered endlessly in a labyrinth of Airy functions without coning to grips with the minotaur,
  the mysterious and peculiar $8/5\sqrt{3}$. Why \emph{is} the polarization for electrons so large, and yet not complete? I have no compelling answer.
\end{quote}
It was after I read the astrophysics papers that I realized the answer to Jackson's challenge.
Here is my explanation.
For electrons orbiting a neutron star, the electron orbits decay as photons are emitted, and the electrons eventually drop to the ground state, which is non-degenerate
(because of the magnetic field).
Hence their asymptotic spin polarization level is indeed $100\%$.
However, for electrons circulating in a storage ring, 
the ring contains radio-frequency cavities which replenish the energy loss of the electrons due to the radiation of photons.
An electron circulating in a storage ring can radiate indefinitely, and its orbit does not decay.
An electron emitting a photon in a storage ring can therefore always find a lower energy state of the opposite spin orientation.
Hence the asymptotic electron spin polarization level in a storage ring is lower than $100\%$.
Jackson kindly accepted my explanation as a satisfactory answer to his challenge \cite{JDJprivcomm}.

\begin{figure}[!htb]
\centering
\includegraphics[width=0.80\textwidth]{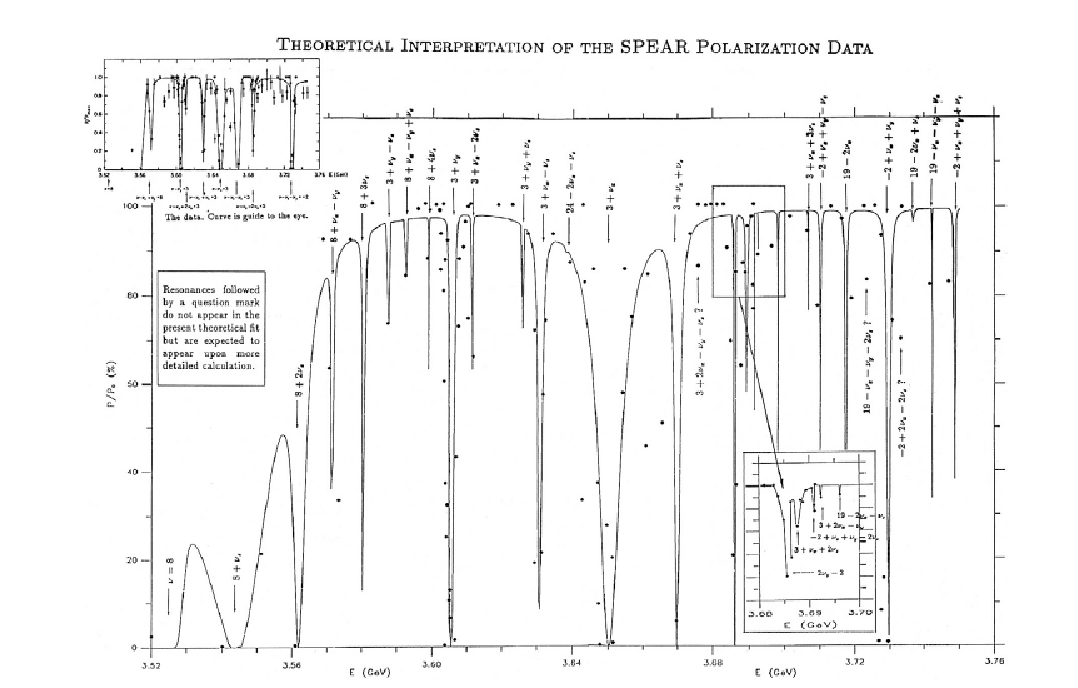}
\caption[\;Theoretical fit of the SPEAR polarization data]{
  \label{fig:SPEAR_fit}
  Theoretical fit of the SPEAR polarization data. From \cite{ManeSPEARfit}.}
\end{figure}

\cite{DK73} extended the theory and published the definitive formula for radiative spin polarization.
The Derbenev-Kondratenko formula treats inhomogeneous fields, for example the focusing fields of quadrupole magnets, etc.,
and an ensemble of electrons with a distribution of orbits in phase-space.
We shall see below that the derivation of their formula contains a much richer set of physics ideas.
In particular, the Derbenev-Kondratenko formula describes the so-called ``depolarizing spin resonances'' which have been observed in storage rings.
Such spin resonances do not appear in the simple model of circular motion in a uniform vertical magnetic field.
\cite{ManePAC89} employed the Derbenev-Kondratenko formula to fit measurements of the electron spin polarization at the SPEAR storage ring \cite{SPEARpol},
which exhibited numerous depolarizing spin resonances.
See Fig.~\ref{fig:SPEAR_fit}.
A major task of this article is to elucidate the formalism pioneered by Derbenev and Kondratenko.

\section{Basic notions}\label{ch:basic}
Throughout this article,
we treat a particle of mass $m$, charge $e$, velocity $\bm{v}=\bm{\beta}c$, Lorentz factor $\gamma=1/\sqrt{1-\beta^2}$ and spin $\bm{s}$.
The particle's coordinate vector is $\bm{r}$, its canonical momentum is $\bm{p}$ and its energy is $E$.
The magnetic moment anomaly is denoted by $a=(g-2)/2$.
We treat spin $\frac12$ particles only.
In that case, the polarization density matrix $\rho_{\rm pol}$ is a $2\times2$ Hermitian matrix with unit trace,
and is specified completely by a three-component polarization vector $\bm{P}$ as follows.
\begin{equation}
\label{eq:polmtx}
\rho_{\rm pol} = \frac12(I + \bm{\sigma}\cdot\bm{P}) = \frac12\begin{pmatrix} 1+P_3 & P_1-iP_2 \\ P_1+iP_2 & 1-P_3\end{pmatrix} \,.
\end{equation}
Here $I$ is the $2\times2$ identity matrix and $\bm{\sigma}$ is a vector of Pauli matrices.
Eq.~\eqref{eq:polmtx} is valid in any coordinate basis.
In the rest of this article, we shall represent the polarization by a three-component vector $\bm{P}$.

\emph{It is important not to confuse the polarization vector $\bm{P}$ with the spin $\bm{s}$.}
They are both three-component vectors but have different meanings.
The spin $\bm{s}$ is a quantum operator and belongs to an individual electron.
The polarization is not a quantum operator.
The polarization is the \emph{expectation value} of the spin distribution.
The polarization is a property of the ensemble of all the particles and is not attached to an individual electron,
or to an individual point in phase-space.
In a storage ring, the polarization is the expectation value of the spin distribution over the occupied phase-space of the particle orbits.
\emph{It is possible for an individual electron to be unpolarized, for example a single electron in a Penning trap.}

In the model we treat, the motion of the individual electrons is independent and the emission of distinct photons is incoherent.
Hence we treat spontaneous photon emission only.
We shall not treat radiation from so-called ``free electron lasers'' and stimulated photon emission. 
This is a standard practice in the literature.
If the particles are initially unpolarized, the polarization level, say $P(t)$,
relaxes exponentially to an equilibrium (or asymptotic) polarization level, say $P_{eq}$,
with a time constant (the ``polarization buildup time''), say $\tau_{pol}$, as given by the following formula
\begin{equation}
\label{eq:Pt}
P(t) = P_{eq}(1 - e^{-t/\tau_{pol}}) \,.
\end{equation}
Let the populations of the ``up'' and ``down'' spin states be $N_\uparrow$ and $N_\downarrow$, respectively.
Let $p_\uparrow$ and $p_\downarrow$ denote respectively the spin-flip probabilities per unit time for flips down-to-up and up-to-down.
Then in equilibrium one must have $p_\uparrow N_\downarrow = p_\downarrow N_\uparrow$.
The equilibrium polarization level is given by
\begin{equation}
\label{eq:Peq}
P_{eq} = \frac{N_\uparrow-N_\downarrow}{N_\uparrow+N_\downarrow} = \frac{p_\uparrow-p_\downarrow}{p_\uparrow+p_\downarrow} \,.
\end{equation}
The polarization buildup time is given by
\begin{equation}
\label{eq:tpol}
\tau_{pol}= \frac{1}{p_\uparrow+p_\downarrow} \,.
\end{equation}
Observe that the nonflip photon emissions do not contribute to the above expressions.
We shall discuss the contributions of the nonflip photon emissions in Chapter \ref{ch:DK}.
As we noted in the Introduction, about radio-frequency cavities in a storage ring,
there are background details which \emph{do} make important contributions to the polarization in real storage rings.

We can already make some pertinent observations, without detailed formalism.
\cite{ST} calculated the radiated power (and the resulting spin polarization) for a model
of ultrarelativistic particle motion in a horizontal circle of radius $\rho$ in a uniform vertical magnetic field.
They employed the Dirac equation, but that is not relevant here.
(In later years, \cite{SchwingerTsaiRadpol} did the same, but they employed source theory.)
We follow \cite{ST} and employ the symbol $W$ for the radiated power, to avoid confusion with $P$ for the polarization.
They obtained the following expressions for the radiated power, to $O(\hbar^2)$.
First define a dimensionless parameter $\xi$
and write the classical radiated power $W^{\rm cl}$ as follows
\begin{equation}
  W^{\rm cl} = \frac23\,\frac{e^2c\gamma^4}{\rho^2} \,,\qquad
  \xi = \frac32\,\frac{\hbar\gamma^2}{mc\rho} \,.
\end{equation}
Recall $m$, $e$ and $\gamma$ are respectively the particle mass, charge and Lorentz factor.
Then
\begin{subequations}
\begin{align}
  W_\sigma^{\uparrow\uparrow} &= W^{\rm cl}\biggl\{\frac78 -\xi\biggl(\frac{25\sqrt3}{12}+\zeta\biggr)
  \nonumber\\
  &\qquad\qquad\quad
  +\xi^2\biggl(\frac{335}{18}+\frac{245\sqrt3}{48}\zeta\biggr)+\dots\biggr\} \,,
  \\
  W_\sigma^{\uparrow\downarrow} &= W^{\rm cl}\,\frac{\xi^2}{18} \,,
  \\
  W_\pi^{\uparrow\uparrow} &= W^{\rm cl}\biggl\{\frac18 -\xi\frac{5\sqrt3}{24} +\xi^2\frac{25}{18} +\cdots\biggr\} \,,
  \\
  W_\pi^{\uparrow\downarrow} &= W^{\rm cl}\,\xi^2\frac{23}{18}\biggl\{1+\zeta\frac{105\sqrt3}{184}\biggr\} \,.
\end{align}
\end{subequations}
The arrows show the relative direction of the spin in the initial and final spin states.
The initial spin state is specified by $\zeta = \pm1$, if the initial spin is directed along/against the field.
The subscripts $\sigma$ and $\pi$ denote the emitted photon polarization, in the orbital plane ($\sigma$ mode) or perpendicular ($\pi$ mode).
It is known, for example, that classical synchrotron radiation is strongly linearly polarized in the orbital plane ($\sigma$ mode).
The spin-flip radiation, on the other hand, is strongly linearly polarized in the vertical direction ($\pi$ mode).

In passing, let us note the following observation.
It is known that the radiated power is a Lorentz invariant.
\cite{Schwinger1949} proved this fact for classical synchrotron radiation.
More generally, \emph{the radiated power in each spin channel is separately Lorentz invariant.}
Two observers in different reference frames must agree whether a spin flipped or not, when a photon is emitted.

The ratio of the spin-flip radiated power to the classical radiated power is
\begin{equation}
\begin{split}
  \frac{W^{\rm spin-flip}}{W^{\rm cl}} &= \frac{W_\sigma^{\uparrow\downarrow} +W_\pi^{\uparrow\downarrow}}{W^{\rm cl}}
  \\
  &= \frac43\xi^2\biggl(1+\zeta\frac{35\sqrt3}{64}\biggr) \,.
\end{split}
\end{equation}
Using $\rho = 1000$ m and $E = 30$ GeV yields $\xi\simeq 9\times 10^{-7}$, and so the ratio of spin-flip to classical radiated power is roughly $5\times 10^{-12}$.
This exemplifies the claim that spin-flip photon emissions are rare, compared with the emission of ordinary synchrotron radiation.
As a practical matter, the spin-flip radiated power is too small to observe with present-day detectors,
but the spin-flip photon emissions leave an imprint on the electron spins, and the electron spin polarization can and has been measured.
\cite{ST} obtained the following expressions for the asymptotic polarization level the polarization buildup time.
\begin{subequations}
\begin{align}
\label{eq:PST}  
  P_{\rm ST} &= \frac{8}{5\sqrt3} \;\simeq 92.376\%\,,
  \\
\label{eq:tauST}  
  \tau_{\rm ST}^{-1} &= \frac{5\sqrt3}{8}\frac{e^2\hbar\gamma^5}{m^2c^2\rho^3} \,.
\end{align}
\end{subequations}
For electrons, the sign is reversed, i.e.~the polarization is aligned against the external magnetic field.
Observe that the asymptotic polarization level is high, more than $90\%$, even though the buildup process is slow.
What matters, for the asymptotic polarization level, is the \emph{asymmetry} of the spin-flip rates in opposite directions, up-to-down vs.~down-to-up.
Spin-flip photon emissions are rare, and the buildup of radiative spin polarization is a slow process.
The Sokolov-Ternov polarization buildup time at HERA was about 43 minutes for a lepton beam energy of $26.7$ GeV \cite{BarberHERApol}.
As can also be seen from eq.~\eqref{eq:tauST}, the Sokolov-Ternov polarization buildup time is also strongly energy-dependent.
At LEP, it was only about seven minutes at a beam energy of 100 GeV but about 2 hours at 55 GeV \cite{Bovet1985}.

If the particle orbit is planar but not circular, but is (for example) a set of circular arcs joined by straight lines,
define an azimuthal angle $\theta$ along the arc-length, so $0\le\theta\le2\pi$ around the ring circumference.
Then the polarization buildup time is given by
\begin{equation}
  \tau^{-1} = \frac{5\sqrt3}{8}\frac{e^2\hbar\gamma^5}{m^2c^2} \int_0^{2\pi} \frac{1}{|\rho|^3}\,\frac{d\theta}{2\pi}\,.
\end{equation}
The absolute value signs on $|\rho|$ are to allow the possibility that the orbit could contain regions of negative curvature.
The asymptotic polarization level is
\begin{equation}
  P_{eq} = P_{\rm ST} \;\frac{\displaystyle \int_0^{2\pi} \frac{1}{\rho^3}\,\frac{d\theta}{2\pi}}{\displaystyle \int_0^{2\pi} \frac{1}{|\rho|^3}\,\frac{d\theta}{2\pi}} \,.
\end{equation}
Observe that this could be lower than $P_{\rm ST}$ if the orbit has regions of negative curvature.
Possibly zero if the ring is a figure-of-eight, although in practice no such ring has ever been built.

The Sokolov-Ternov calculation was for ultratelativistic electrons.
Observe that the asymptotic level $P_{\rm ST}=8/(5\sqrt3)$ is independent of $\gamma$, but the value of $\tau_{\rm ST}$ depends strongly on $\gamma$.  
Clearly the value of $P_{\rm ST}$ is a limiting value, for $\gamma\to\infty$.
\cite{JacksonRMP} referred to it as the ``minotaur'' (see the quote from Jackson in the Introduction).
I set myself the challenge to calculate the asymptotic polarization level as a function of $\beta$, for $\beta\ll1$.
I obtained the following result
\begin{equation}
P_{eq} \simeq 1 - \frac{3}{28}\beta^4 \,.
\end{equation}
One expects the result must be a function of $\beta^2$, because the polarization level must be the same for either sense of circulation around the ring.
However, it surprised me that the term in $\beta^2$ was zero.
Note also that as $\beta\to0$ the asymptotic polarization level does indeed approach unity.
More interesting is to take the limit $\beta\to1$. One obtains 
\begin{equation}
P_{eq} \to 1 - \frac{3}{28} = \frac{25}{28} \simeq 89.286\% \,.
\end{equation}
Subtracting from the Sokolov-Ternov value, the difference is only about $3\%$:
\begin{equation}
\Delta P_{eq} = \frac{8}{5\sqrt3} -\frac{25}{28} \simeq 3.09\% \,.
\end{equation}
\emph{All of the complications of relativity add only an extra $3\%$ to the nonrelativistic calculation.}
Of course, the polarization buildup time is enormous: relativity contributes a factor of $\gamma^5$, which cannot be neglected.

\begin{figure}[!ht]
\centering
\includegraphics[width=0.45\textwidth]{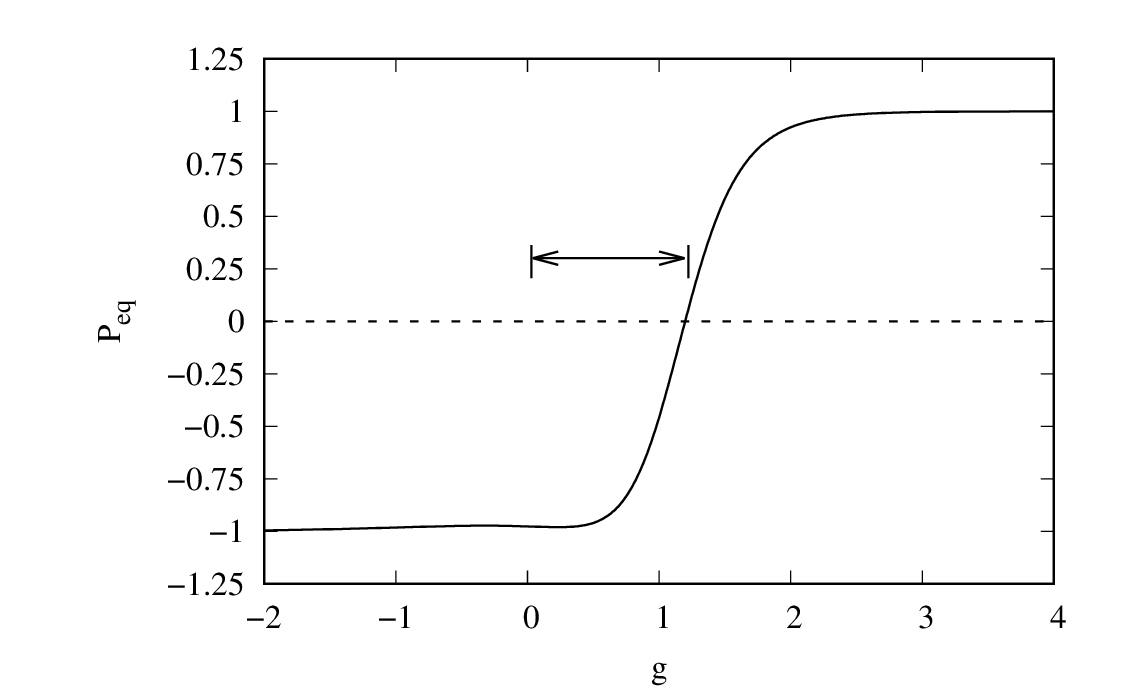}
\caption[\;Graph of asymptotic polarization vs.~$g$]{
\label{fig:graph_pol}
Graph of the asymptotic polarization $P_{eq}$ as a function of $g$, for a model of circular motion in a uniform vertical magnetic field.
For $0 < g \le 1.2$, the polarization is negative, i.e.~the na{\i}vely higher-energy spin state is preferentially populated.
The range is indicated by the horizontal arrow.}
\end{figure}

\begin{figure}[!hb]
\centering
\includegraphics[width=0.45\textwidth]{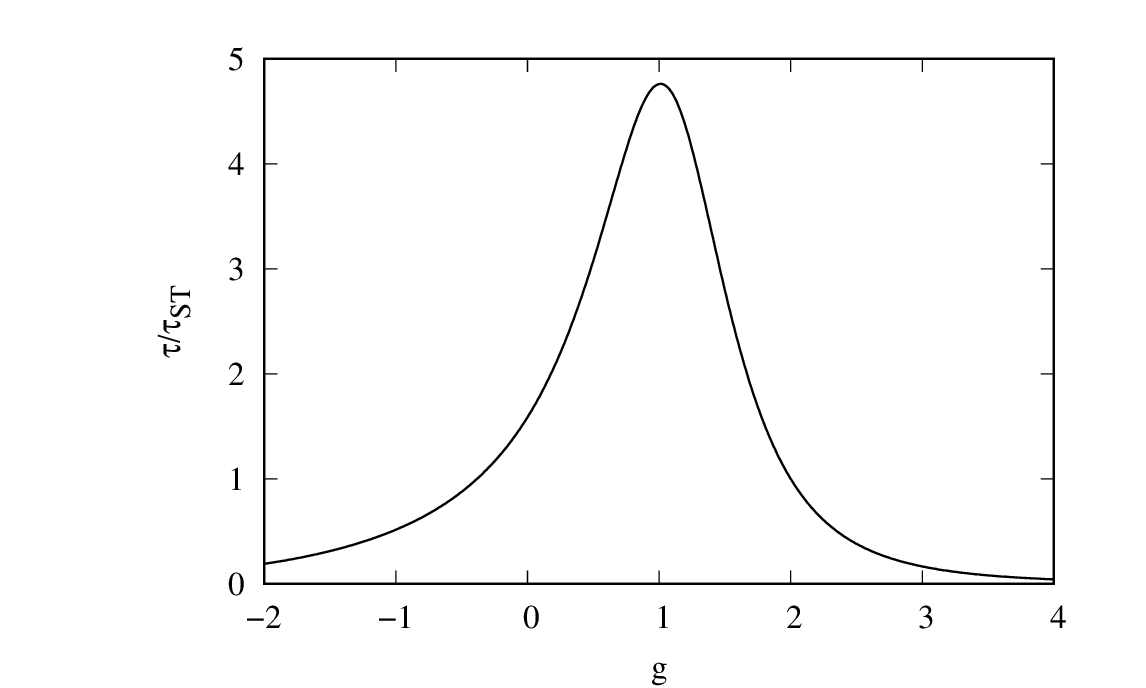}
\caption[\;Graph of polarization buildup time vs.~$g$]{
\label{fig:graph_tau}
Graph of the polarization buildup time $\tau$ (actually the ratio $\tau/\tau_{\rm ST}$) as a function of $g$, for a model of circular motion in a uniform vertical magnetic field.}
\end{figure}

Because \cite{ST} employed the Dirac equation, their calculation was for the case $g = 2$.
In fact, the values of the asymptotic polarization and the buildup time depend on $g$.
The generalization to arbitrary $g$ was given by \cite{DK73}, using semiclassical QED.
Observe that the value of $a=(g-2)/2$ in eq.~\eqref{eq:H0} is a free parameter.
The expressions below follow \cite{JacksonRMP}:
\begin{equation}
\begin{split}
  F_1(a) &= 1 +\frac{41}{45}a -\frac{23}{18}a^2 -\frac{8}{15}a^3 +\frac{14}{15}a^4
  \\
  &\quad
  -\frac{8}{5\sqrt3}\frac{a}{|a|}\Bigl(1 +\frac{11}{12}a -\frac{17}{12}a^2 -\frac{13}{24}a^3 +a^4\Bigr) \,,
\\
F_2(a) &= \frac{8}{5\sqrt3}\Bigl(1 +\frac{14}{3}a +8a^2
\\
&\qquad\qquad\quad
+\frac{23}{3}a^3 +\frac{10}{3}a^4 +\frac{2}{3}a^5\Bigr) \,.
\end{split}
\end{equation}
The asymptotic polarization level and buildup time are
\begin{subequations}
\begin{align}
  P_{eq}(a) &= \frac{F_2(a)}{\displaystyle F_1(a)e^{-\sqrt{12}|a|} +\frac{a}{|a|}F_2(a)} \,,
  \\
  \tau(a) &= \frac{\tau_{ST}}{\displaystyle F_1(a)e^{-\sqrt{12}|a|} +\frac{a}{|a|}F_2(a)} \,.
\end{align}
\end{subequations}
A graph of the asymptotic polarization $P_{eq}(a)$ is plotted as a function of $g$ in Fig.~\ref{fig:graph_pol}
and a graph of $\tau(a)/\tau_{\rm ST}$ is plotted as a function of $g$ in Fig.~\ref{fig:graph_tau}.
Figure \ref{fig:graph_pol} especially contains some notable features.
\begin{enumerate}
\item
  As expected, the asymptotic polarization level approaches $\pm1$ as $g\to\pm\infty$.
  Essentially, the magnetic dipole moment becomes infinitely large in these limits.
\item
  \cite{DK73} noted that the polarization is \emph{not} zero if $g=0$, i.e.~zero magnetic moment.
  The polarization is then generated purely by the Thomas precession.
  The asymptotic value is \emph{negative}, approximately $-98\%$, and is \emph{higher} (in absolute value) than the Sokolov-Ternov value of $92.4\%$ for $g=2$.
\item
  There is an interval of values $0 < g \lesssim 1.2$, demarcated by the arrow in Fig.~\ref{fig:graph_pol},
  where $g$ is positive but the asymptotic polarization level is \emph{negative}, i.e.~the na{\"\i}vely higher-energy spin state is preferentially populated.
  This shows that the contribution of the Thomas precession is significant.
  The coupling of the electron spins to the radiation field is not purely a magnetic dipole moment interaction.  
\end{enumerate}
I also sent Jackson a (very approximate) perturbation-theoretic calculation of the asymptotic polarization for $|\beta|\ll1$ and $|g|\ll1$.
I found that the polarization is negative for $0<g<2\sqrt{2/5}\beta$.
Taking the limit $\beta\to1$ yields $g_* = 2\sqrt{2/5}\simeq 1.26$.
This is to be compared to the actual value $g_* \simeq 1.2$ in Fig.~\ref{fig:graph_pol}.
My analysis was very approximate, but a nonrelativistic calculation \emph{did} reveal that the asymptotic polarization is negative for some values $g>0$.
It is a competition between two perturbation theories, small $|\beta|$ and small $|g|$.

We briefly discuss the orbital motion in actual storage rings.
The reference for ``strong-focusing synchrotrons'' is the classic paper by \cite{CourantSnyder}.
A more modern and comprehensive textbook is by \cite{EdwardsSyphers}.
For a circulator accelerator such as a storage ring, the particle orbits consist of oscillations around a central orbit called the \emph{closed orbit}.
As its name implies, the closed orbit closes on itself after one circuit around the ring, i.e.~it is a periodic loop around the ring circumference.
\cite{CourantSnyder} proved that the closed orbit always exists, for a planar ring with only magnetic fields.
As a practical matter, particle accelerators are not built with arbitrary topologies: real rings are constructed such that a closed orbit \emph{does} exist.
In passing, note the difference between the \emph{design orbit} and the \emph{closed orbit}.
The design orbit is typically a set of circular arcs joined by straight lines, i.e.~an idealized loop.
Because of inevitable tolerances in the manufacture and misalignment of the magnets in a storage ring,
the actual closed orbit is distorted and does not exactly equal the design orbit.
By the implicit function theorem, if the tolerances in manufacture and misalignments are sufficiently small, the closed orbit will exist.

The other orbits oscillate around the closed orbit and are not periodic around the ring circumference.
The transverse orbital oscillations are easy to visualize.
For historical reasons, they are called \emph{betatron oscillations}.
There are also longitudinal oscillations, which, also for historical reasons, are called \emph{synchrotron oscillations}.
A simplified (oversimplified?) explanation of the longitudinal oscillations is as follows.
It was stated earlier that storage rings contain radio-frequency cavities to replenish the energy loss due to synchrotron radiation.
(There is a reference energy, which is the operating energy of the storage ring.)
However the radio-frequency cavities do more than that; they also provide longitudinal focusing of the particle orbits.
The electrons have a distribution of energies, or longitudinal momenta, hence their times of arrival at a radio-frequency cavity are not all the same.
A radio-frequency cavity contains an oscillating longitudinal electric field, which imparts a longitudinal push to the momentum of a particle.
The particles receive (slightly) different pushes depending on their time of arrival.
The phasing of the ``differential push'' (the push relative to that for a particle on the closed orbit)
is chosen to generate longitudinal focusing of the particle orbits.
Hence an electron ``bunch'' has a longitudinal as well as a transverse extent.
\emph{Note: radio-frequency cavities also provide longitudinal focusing in proton rings, i.e.~no synchrotron radiation.
The emission of synchrotron radiation is not relevant for the longitudinal focusing of the particle orbits.}

It is conventional in accelerator physics to measure the orbital oscillation frequencies in units of the orbital revolution frequency.
The resulting dimensionless oscillation frequency is called a \emph{tune},
viz.~the horizontal and vertical betatron tunes and the synchrotron tune.
The orbital tunes are denoted by $(\nu_1,\nu_2,\nu_3)$.
It is also a common notation to write $(Q_1,Q_2,Q_3)$ instead.
We shall also employ the notation ``$Q$'' below, for the tune of a generic orbital oscillation mode.

The spin tune is the spin precession frequency divided by the orbital revolution frequency.
By convention, the spin tune is denoted by $\nu$, without a subscript.
A so-called ``spin resonance'' occurs when the spin precession resonates with the orbital oscillations.
The formula (or resonant condition) for a spin resonance is
\begin{equation}
\label{eq:spinresformula}
\nu +r_1Q_1 +r_2Q_2 +r_3Q_3 = r \,.
\end{equation}
Here $r_1$, $r_2$, $r_3$ and $r$ are integers (not necessarily positive).
Typically, the polarization drops to zero at a spin resonance.
Hence many authors call them ``depolarizing spin resonances'' and we have done so ourselves.
See the labelling of the spin resonances in Fig.~\ref{fig:SPEAR_fit}, and observe that the polarization is zero at the center of a spin resonance.
Note also the following:
\begin{enumerate}
\item
The coefficient of the spin tune in eq.~\eqref{eq:spinresformula} is always unity. For particles with spin $\frac12$,
any product of Pauli matrices always reduces to a linear combination of the identity matrix and Pauli matrices.
For the orbit, however, it is possible for multiple orbital oscillations to compound with each other in the spin-orbit coupling.
\item
Orbital resonances, not involving the spin, occur whenever a linear combination of the orbital tunes sums to an integer,
i.e.~$r_1Q_1 +r_2Q_2 +r_3Q_3 = r$.
The orbital motion is unstable at an orbital resonance.
Throughout this article, we assume the orbital motion is stable.
\end{enumerate}

\section{Storage rings \&\ Derbenev-Kondratenko formula}\label{ch:DK}
\subsection{Semiclassical Hamiltonian}\label{sec:Ham}
Clearly, it is too complicated to treat real storage rings using the Dirac equation.
Subsequent authors after Sokolov and Ternov showed a semiclassical Hamiltonian is adequate.
For highly relativistic charged particles, the Stern-Gerlach force is negligible.
As a practical matter, it has long been known that the orbit of a particle in a high-energy accelerator (either linear or circular)
does not depend on the particle's spin orientation.
However, the spin precession \emph{does} depend on the particles's orbit:
for example if the particle passes off-axis through a quadrupole magnet, etc.
Hence the orbital motion can be treated classically, but the spin precession depends on the orbit.
Hence in this context, ``semiclassical'' means the orbital motion is treated classically.
The spin is a two-valued quantum operator.

Since $\bm{s}\cdot\bm{s}$ is a constant, the most general equation of motion for the spin is a precession equation
\begin{equation}
\label{eq:spineq}
\frac{d\bm{s}}{dt} = \bm{\Omega}\times\bm{s} \,.
\end{equation}
Here $\bm{\Omega}$ is the spin precession vector.
In the case where the spin precesses in prescribed external electromagnetic fields $\bm{E}_{\rm ext}$ and $\bm{B}_{\rm ext}$,
eq.~\eqref{eq:spineq} is called the Thomas-BMT equation \cite{Thomas}, \cite{BMT} and $\bm{\Omega}$ is given by
\begin{equation}
\label{eq:Omegaext}
\begin{split}
\bm{\Omega}_{\rm ext} &= -\frac{e}{mc}\biggl[\biggl(a+\frac{1}{\gamma}\biggr)\bm{B}_{\rm ext} -\frac{a\gamma}{\gamma+1}(\bm{\beta}\cdot\bm{B}_{\rm ext})\bm{\beta}
\\
&\qquad\qquad\quad
  -\biggl(a+\frac{1}{\gamma+1}\biggr)\bm{\beta}\times\bm{E}_{\rm ext}\biggr] \,.
\end{split}
\end{equation}
\emph{Note the following important detail, which has not been stated explicitly up to now:}
$\bm{s}$ is the spin operator in the particle's \emph{rest frame},
whereas the particle coordinate, momentum and the electromagnetic fields and potentials are all in the \emph{laboratory frame}.
Hence $\bm{s}$ does \emph{not} change under a Lorentz boost.

The semiclassical Hamiltonian for the orbital and spin motion in the storage ring is
\begin{equation}
\label{eq:H0}
\begin{split}
  H_0 &= \biggl[\biggl(\bm{p} -\frac{e}{c}\bm{A}_{\rm ext}\biggr)^2 +m^2c^4\,\biggr]^{1/2}
  \\
  &\qquad +e\Phi_{\rm ext} +\bm{\Omega}_{\rm ext}\cdot\bm{s} \,.
\end{split}
\end{equation}
Here $\Phi_{\rm ext}$ and $\bm{A}_{\rm ext}$ are respectively the scalar and vector potentials of the prescribed external electric and magnetic fields of the storage ring.
Our perturbation theory is QED (quantum electrodynamics).
The unperturbed Hamiltonian $H_0$ describes the electron motion in the prescribed fields of the storage ring.
The interaction Hamiltonian $H_{\rm int}$ describes the electron interaction with the photon (radiation) EM fields
\begin{equation}
\label{eq:Hint}
H_{\rm int} = -e\bm{\beta}\cdot\bm{A}_{\rm rad} +\bm{\Omega}_{\rm rad}\cdot\bm{s} \,.
\end{equation}
Here $\bm{A}_{\rm rad}$ is the vector potential of the photon (radiation) field.
We employ the radiation gauge, where $\bm{\nabla}\cdot\bm{A}_{\rm rad}=0$ and the scalar potential $\Phi_{\rm rad}$ of the radiation field is zero.
Also $\bm{\Omega}_{\rm rad}$ has the same form as $\bm{\Omega}_{\rm ext}$ but with radiation EM fields substituted for the prescribed external EM fields.
By definition, $H_{\rm int}$ is Hermitian, hence it contains terms for both photon emission and absorption.
We are concerned only with photon emission.

\subsection{Spin quantization axis}\label{sec:DKnaxis}
We require a spin projection operator which is part of a complete set of commuting observables.
Following \cite{DK73}, we seek a unit vector, call it $\bm{n}$, such that $\bm{s}\cdot\bm{n}$ is a dynamical invariant.
Then $\bm{n}$ must itself be a dynamical variable, i.e.~a function of the orbital phase-space variables $(\bm{r},\bm{p})$ only.
This is expressed by stating that $\bm{n}$ is not an explicit function of the time:
\begin{equation}
\label{eq:npartial}
\frac{\partial\bm{n}}{\partial t} = 0 \,.
\end{equation}
Next, we demand that $\bm{n}$ satisfy the spin precession equation eq.~\eqref{eq:spineq}.
\begin{equation}
\label{eq:dndt}
\frac{d\bm{n}}{dt} = \bm{\Omega}\times\bm{n} \,.
\end{equation}
Then the spin projection operator $\bm{s}\cdot\bm{n}$ is a dynamical variable and is also invariant:
\begin{equation}
\begin{split}
  \frac{d(\bm{s}\cdot\bm{n})}{dt} &= \frac{d\bm{s}}{dt}\cdot\bm{n} +\bm{s}\cdot\frac{d\bm{n}}{dt}
  \\
  &= (\bm{\Omega}\times\bm{s})\cdot\bm{n} +\bm{s}\cdot(\bm{\Omega}\times\bm{n})
  \\
  &= 0 \,.
\end{split}
\end{equation}
Eqs.~\eqref{eq:npartial} and \eqref{eq:dndt} together furnish the conditions to specify the spin quantization axis $\bm{n}$.
Observe that the spin projection of $\bm{s}$ along $\bm{\Omega}$ is not an invariant, if $\bm{\Omega}$ is not constant:
Let $\hat{\bm{\Omega}}$ be the unit vector along $\bm{\Omega}$, then
\begin{equation}
\begin{split}
  \frac{d(\bm{s}\cdot\hat{\bm{\Omega}})}{dt} &= \frac{d\bm{s}}{dt}\cdot\hat{\bm{\Omega}} +\bm{s}\cdot\frac{d\hat{\bm{\Omega}}}{dt}
  \\
  &= (\bm{\Omega}\times\bm{s})\cdot\hat{\bm{\Omega}} +\bm{s}\cdot\frac{d\hat{\bm{\Omega}}}{dt}
  \\
  &= \bm{s}\cdot\frac{d\hat{\bm{\Omega}}}{dt}
  \\
  &\ne 0 \,.
\end{split}
\end{equation}
However, if $\bm{\Omega}$ is constant, then $d\hat{\bm{\Omega}}/dt=0$.
In this special case, e.g.~a uniform vertical magnetic field, then $\bm{n}=\hat{\bm{\Omega}}$.

\cite{DK73} proved, with some caveats, that $\bm{n}$ is unique, up to a minus sign.
The proof is as follows.
Suppose there are two distinct solutions for $\bm{n}$, say $\bm{n}_1$ and $\bm{n}_2$ and they are not parallel, so $\bm{n}_1\times\bm{n}_2\ne0$.
Then define $\bm{n}_3 = \bm{n}_1\times\bm{n}_2$. (This is a slight abuse of notation, because $\bm{n}_3$ is not a unit vector.)
Then clearly $\partial\bm{n}_3/\partial t=0$. Next, skipping some intermediate steps in the algebra for brevity,
\begin{equation}
\begin{split}
  \frac{d\bm{n}_3}{dt} &= \frac{d\bm{n}_1}{dt}\times\bm{n}_2 +\bm{n}_1\times\frac{d\bm{n}_2}{dt}
  \\
  &= (\bm{\Omega}\times\bm{n}_1)\times\bm{n}_2 +\bm{n}_1\times(\bm{\Omega}\times\bm{n}_2)
  \\
  &= \bm{\Omega}\times(\bm{n}_1\times\bm{n}_2)
  \\
  &= \bm{\Omega}\times\bm{n}_3 \,.
\end{split}
\end{equation}
In this case, $(\bm{n}_1,\bm{n}_2,\bm{n}_3)$ form a linearly independent triad, hence \emph{every} unit vector satisfies
eqs.~\eqref{eq:npartial} and \eqref{eq:dndt}.
This happens at the spin resonances mentioned in Chapter \ref{ch:basic}:
$\bm{n}$ is ill-defined at a spin resonance.
Technically, \cite{DK73} proved that \emph{if} $\bm{n}$ exists, \emph{then} it is unique up to a minus sign.
No one has proved that $\bm{n}$ always exists, but as a practical matter storage rings are not built with arbitrary topologies:
real rings are constructed such that $\bm{n}$ \emph{does} exist (except at a spin resonance).

\subsection{Spin-flip photon emission matrix element}\label{sec:DKspinflip}
The fact that $\bm{n}$ is not a single axis but is a function of the orbital phase-space location has profound implications.
(Stated more formally, $\bm{n}$ is a unit vector field over the manifold of the phase-space of the particle orbits.)
When a photon is emitted, the electron recoils to a \emph{different} point in the orbital phase-space.
This was the great insight of \cite{DK73}:
spin-flip in a real storage ring is \emph{not} a $180^\circ$ reversal of the spin orientation.
Spin-flip is a transition from a spin state quantized along an initial axis $\bm{n}_i$ before emission, say $|\pm\bm{n}_i\rangle$
to a final spin state $|\mp\bm{n}_f\rangle$ quantized along a \emph{different} final axis $\bm{n}_f$ after emission.
The spin-flip photon emission matrix element is then as follows.
\begin{equation}
\label{eq:mtxel}  
\begin{split}
  \langle\mp\bm{n}_f|H_{\rm int}|\pm\bm{n}_i\rangle &= \langle\mp\bm{n}_f|\bigl(-e\bm{\beta}\cdot\bm{A}_{\rm rad}
  \\
  &\qquad\qquad
  +\bm{\Omega}_{\rm rad}\cdot\bm{s}\bigr)|\pm\bm{n}_i\rangle 
  \\
  &= -e\bm{\beta}\cdot\bm{A}_{\rm rad} \underbrace{\langle\mp\bm{n}_f|\pm\bm{n}_i\rangle}_{\ne0}
  \\
  &\quad + \bm{\Omega}_{\rm rad}\cdot\langle\mp\bm{n}_f|\bm{s}|\pm\bm{n}_i\rangle \,.
\end{split}
\end{equation}
The consequences of the insight of \cite{DK73} are profound:
\emph{even classical synchrotron radiation can flip the spin.}
The second term in eq.~\eqref{eq:mtxel} was always obvious: the spin operator $\bm{s}$ couples spin states of opposite orientation.
In the first term in eq.~\eqref{eq:mtxel}, $-e\bm{\beta}\cdot\bm{A}_{\rm rad}$ is the term responsible for classical synchrotron radiation.
This term in $H_{\rm int}$ is spin-independent.
However, it generates a recoil of the electron momentum, and that changes the electron's phase-space location,
and $\bm{n}_i$ and $\bm{n}_f$ are not parallel (in general), hence the overlap $\langle\mp\bm{n}_f|\pm\bm{n}_i\rangle$ is nonzero.
This is an indirect mechanism of flipping the spin, but remember that $-e\bm{\beta}\cdot\bm{A}_{\rm rad}$ is of order unity,
whereas $\bm{s}$ in the second term in $H_{\rm int}$ is $O(\hbar)$.
    
For the idealized model ring of a uniform magnetic field, $\bm{n}$ is always vertical, hence the first term in eq.~\eqref{eq:mtxel} is absent.
However, as we shall see later, it is this term which drives the ``depolarizing spin resonances'' and, at sufficiently high energies,
it is stronger than the second term in eq.~\eqref{eq:mtxel}.

\subsection{Derbenev-Kondratenko formula}\label{sec:DKformula}
To evaluate the spin-flip matrix element in eq.~\eqref{eq:mtxel}, we proceed as follows.
We first note that, although $\bm{n}_i$ and $\bm{n}_f$ are not parallel, they are \emph{almost} parallel.
The energy of an emitted photon is much less than the energy of the electron, hence the change in the electron's phase-space location is small.
Furthermore, it is well-known that the direction of photon emission is almost longitudinal, to $O(1/\gamma)$ (``relativistic beaming'').
Hence it is a good approximation to treat the electron recoil as parallel to its momentum.
Furthermore, for ultrarelativistic electrons, $\Delta p \simeq \Delta E/c$,
hence we approximate the change in the electron's phase-space location as a loss of energy (a scalar as opposed to a vector derivative).
Then, writing $\Delta\bm{n}=\bm{n}_f-\bm{n}_i$, we obtain
\begin{equation}
\begin{split}
  \Delta\bm{n} &\simeq \Delta E\,\frac{\partial \bm{n}}{\partial E}
  \\
  &= \frac{\Delta E}{E}\,\biggl(\gamma\frac{\partial\bm{n}}{d\gamma}\biggr)
  \\
  &= -\frac{\hbar\omega_{\rm photon}}{E}\,\biggl(\gamma\frac{\partial\bm{n}}{d\gamma}\biggr) \,.
\end{split}
\end{equation}
This introduces the phase-space derivative $\gamma(\partial\bm{n}/d\gamma)$, which will play a key role below.

We can now proceed in one of two directions.
\cite{DK73} attached $\bm{n}$ to an individual particle and followed it around in the orbital phase-space, and took suitable
time averages over the orbital and spin motion.
\cite{Manederiv} calculated spin-flip transition probabilities per unit time, etc., in an infinitesimal phase-space volume element
and took suitable averages over the orbital and spin phase-spaces.
By the hypothesis that the particle motion is ergodic, the two formalisms yield the same answer.
The result for the asymptotic polarization is the Derbenev-Kondratenko formula
\begin{equation}
\label{eq:PDK}
  P_{\rm DK} = \frac{8}{5\sqrt3}\,
  \frac{\displaystyle \biggl\langle\frac{1}{|\rho|^3}\,\hat{\bm{b}}\cdot\biggl(\bm{n} -\gamma\frac{\partial\bm{n}}{d\gamma}\biggr)\biggr\rangle}
    {\displaystyle \biggl\langle\frac{1}{|\rho|^3}\biggl(1 -\frac29(\bm{n}\cdot\hat{\bm{v}})^2
      +\frac{11}{18}\biggl|\gamma\frac{\partial\bm{n}}{d\gamma}\biggr|^2 \biggr)\biggr\rangle} \,.
\end{equation}
Here $\bm{v}$ is the electron velocity,
$\hat{\bm{b}}$ is a unit vector along $\bm{v}\times\dot{\bm{v}}$, essentially a unit vector in the direction of the local magnetic field,
and $\rho$ is the local bend radius of the particle orbit.
The angle brackets denote an integral around the ring circumference and an average over the phase-space of particle orbits.
The corresponding polarization buildup time is
\begin{equation}
  \label{eq:tauDK}
  \tau_{\rm DK} = \tau_{\rm ST}\,\frac{\displaystyle\biggl\langle\frac{1}{|\rho|^3}\biggr\rangle}{\displaystyle
  \biggl\langle\frac{1}{|\rho|^3}\biggl(1 -\frac29(\bm{n}\cdot\hat{\bm{v}})^2 +\frac{11}{18}\biggl|\gamma\frac{\partial\bm{n}}{d\gamma}\biggr|^2 \biggr)\biggr\rangle} \,.
\end{equation}
The Derbenev-Kondratenko formula eq.~\eqref{eq:PDK} and the associated buildup time eq.~\eqref{eq:tauDK} are the key formulas in the theory of radiative spin polarization.

There is important physics implicit in the phase-space averages in eqs.~\eqref{eq:PDK} and \eqref{eq:tauDK}.
First, classical synchrotron radiation is simply the expectation value of the emission of (many) discrete photons.
There are fluctuations in the photon emissions, which cause the electrons to undergo random recoils.
Those random recoils in turn cause the electrons to execute random walks in the orbital phase-space.
The orbital motion of the electrons settles down to an equilibrium distribution in the orbital phase-space.
The equilibration time of the orbital motion is of the order of tens to hundreds on milliseconds in present-day storage rings.
This is \emph{much} faster than the equilibration time of the polarization, which is several minutes to hours,
i.e.~$\tau_{\rm pol} \gg \tau_{\rm orbit}$.
An electron emits many nonflip photons and traverses the orbital phase-space many times between successive spin-flip photon emissions.
Hence it is a good approximation to assume the orbital distribution of the electrons is in equilibrium, when calculating the polarization.
This approximation is made by all authors in the field.
With a reasonably obvious notation, the polarization vector is given by
\begin{equation}
\bm{P} = \langle\langle\bm{s}\cdot\bm{n}\rangle\,\bm{n}\rangle \,.
\end{equation}
Because $\tau_{\rm pol} \gg \tau_{\rm orbit}$, the value of $\langle\bm{s}\cdot\bm{n}\rangle$ is uniform across the orbital phase-space.
This is assumed so in Derbenev and Kondratenko's formalism.
Hence we can factorize, with a very good aproximation,
\begin{equation}
\bm{P} \simeq \langle\bm{s}\cdot\bm{n}\rangle\,\langle\bm{n}\rangle \,.
\end{equation}
The value of $P_{\rm DK}$ in eq.~\eqref{eq:PDK} is the term
$P_{\rm DK} = \langle\bm{s}\cdot\bm{n}\rangle$.
The factor of $\langle\bm{n}\rangle$ specifies the direction of the polarization vector,
but its magnitude is almost unity and that magnitude is ignored by all workers in the field.
I am indebted to \cite{Yokoyapc} for explaining to me the details of the phase-space averaging implicit in the derivation of the Derbenev-Kondratenko formula.

\subsection{Comments}
We now offer a few observations about eqs.~\eqref{eq:PDK} and \eqref{eq:tauDK}.
\begin{enumerate}
\item
  For a model of motion in a horizontal circle of radius $\rho$ in a uniform vertical magnetic field, $\bm{n}$ is vertical everywhere,
  hence $\gamma(\partial\bm{n}/d\gamma)=0$ and also $\bm{n}\cdot\hat{\bm{v}}=0$.
  Then we recover the Sokolov-Ternov expressions in eqs.\eqref{eq:PST} and \eqref{eq:tauST}.
\item
  The term in $\hat{\bm{b}}\cdot\gamma(\partial\bm{n}/d\gamma)$ in the numerator of eq.~\eqref{eq:PDK}
  comes from the quantum-mechanical interference between the first and second terms in the spin-flip matrix element in eq.~\eqref{eq:mtxel}.
  It is very small and has never been observed in practice.
\item
  The term in $(\bm{n}\cdot\hat{\bm{v}})^2$ in the denominator of eq.~\eqref{eq:PDK} is also negligible.
  It is almost zero in a planar ring because $\bm{n}$ and $\hat{\bm{v}}$ are almost orthogonal.
  In a nonplanar ring such as HERA, the term in $|\gamma(\partial\bm{n}/d\gamma)|^2$ is far more significant.  
\item
  The term in $|\gamma(\partial\bm{n}/d\gamma)|^2$ comes from the (absolute) square of the first term in eq.~\eqref{eq:mtxel}.
  It is important and will be discussed below.
  It is responsible for the depolarizing spin resonances, as displayed in Fig.~\ref{fig:SPEAR_fit}, for example.
\item
  The same denominator expression appears in both the formulas for the asymptotic polarization eq.~\eqref{eq:PDK}
  and the polarization buildup time eq.~\eqref{eq:tauDK}.
  Hence the asymptotic polarization level and the buildup time are correlated:
  as the asymptotic polarization level increases, the buildup time also increases.
  To put it another way, in a real storage ring, a higher degree of asymptotic polarization also implies a slower rate of polarization buildup.
  This is not evident from the Sokolov-Ternov formula, because $P_{\rm ST}$ in eq.\eqref{eq:PST} is a constant and is independent
  of the polarization buildup time in \eqref{eq:tauST}.
\item
  Returning to the spin-flip matrix element, the first term in eq.~\eqref{eq:mtxel},
  viz.~$-e\bm{\beta}\cdot\bm{A}_{\rm rad}\langle\mp\bm{n}_f|\pm\bm{n}_i\rangle$,
  is fundamentally a depolarizing term.
  It has the same magnitude for both directions of spin-flip, up-down or down-up.
  This additional mechanism of spin-flip is depolarizing, in real storage rings.
  It is manifested by the appearance of $|\gamma(\partial\bm{n}/d\gamma)|^2$ in the denominator of eq.~\eqref{eq:PDK}.
  In general, the asymptotic polarization level in actual storage rings is lower than the Sokolov-Ternov value of $8/(5\sqrt3)\simeq92.4\%$.
\end{enumerate}

\subsection{Scaling with energy}
We shall see in Chapter \ref{ch:ISF} that,
to the leading order of approximation, $\gamma(\partial\bm{n}/d\gamma)$ is proportional to $(a\gamma+1)$ times the gradient of the magnetic field $\bm{B}_{\rm ext}$.
(The gradient is zero for a uniform vertical magnetic field and $\gamma(\partial\bm{n}/d\gamma)=0$, as we have noted.)
Since $|\gamma(\partial\bm{n}/d\gamma)|^2$ appears in the denominator of the
Derbenev-Kondratenko formula eq.~\eqref{eq:PDK} and $a\gamma\gg1$ for a beam energy $E$ of several GeV, this means that at
high energies, the asymptotic polarization level scales according to
\begin{equation}
\label{eq:Pscaling}
P_{eq} = \frac{8}{5\sqrt3}\,\frac{1}{1 +\alpha^2E^2} \,.
\end{equation}
Here $\alpha$ is a phenomenological constant.
The attainable degree of radiative polarization decreases as the beam energy increases.
As far as I can trace the matter, this scaling was first pointed out by \cite{Buonscaling}.

Buon's scaling law worked remarkably well for beam energies up to the $Z^0$ resonance (beam energy of $46.5$ GeV).
See Fig.~3 in \cite{Pol-energy-machines}, 
which displays a graph of the asymptotic polarization level attained at multiple high-energy $e^+e^-$ storage rings.
The dashed and solid curves were computed with and without harmonic spin matching, respectively.
Spin matching will be explained in Chapter \ref{ch:ISF} below.
The scaling law in eq.~\eqref{eq:Pscaling} was derived using a first-order approximation to calculate $\gamma(\partial\bm{n}/d\gamma)$.
At sufficiently high beam energies, higher-order terms become significant and the attainable polarization is even lower.
Radiative spin polarization in LEP was observed up to a maximum beam energy of $60.6$ GeV \cite{AssmannRadPol},
but could not be observed at higher beam energies, in the so-called LEP2 phase of operation of LEP (see Fig.~10 in \cite{LEPhistory}).

\section{Unruh effect \&\ accelerating frames}\label{ch:BL}
\cite{Unruh} discovered that an observer in a uniformly (linear) accelerating reference frame
observes that the vacuum electromagnetic fluctuations have a thermal (blackbody) spectrum, with a tempertaure proportional to the proper acceleration.
John Bell and coworkers investigated the problem, to search for ways to detect such an effect experimentally.
For linear acceleration, the attainable acceleration using present-day technology is too small to be useful.
However, the acceleration is much larger for (uniform) circular motion.
Bell and coworkers selected the coupling of the vacuum electromagnetic fluctuations to the magnetic moment of an electron in a circularly accelerating reference frame,
to employ the polarization of the electron spin as a possible detector for the (circular) Unruh effect.
\cite{BL} published a formula for the resulting asymptotic polarization of the electron spin.
They employed a model of an electron orbiting in a horizontal circle in a uniform magnetic field,
with a homogeneous weak electric field to focus the particle motion, a so-called ``weak focusing'' storage ring.
(In their model, the electron was at rest and the reference frame was accelerating.)
\cite{BL} obtained the following formula for the asymptotic polarization
(the expression below is taken from \cite{BM}, eq.~(17))
\begin{equation}
\label{eq:PBL}
P_{BL} = \frac{8}{5\sqrt3}\frac{1 - \frac{f}{6}}{1 -\frac{f}{18} +\frac{13}{360}f^2} \,.
\end{equation}
Here $f$ is given by
\begin{equation}
\label{eq:fBL}
f = \frac{(g-2)Q^2}{Q^2-\nu^2} \,.
\end{equation}
Here $\nu$ is the spin tune and $Q$ is the tune of the vertical betatron oscillations.
The vertical oscillations are induced by the random vertical recoils of the electrons due to the photon emissions.
Observe that $f$ diverges at $\nu=Q$, a depolarizing spin resonance and the asymptotic polarization level in this case iz zero.
When $f\simeq -2$, the asymptotic polarization reaches a maximum value of approximately $98\%$.
This is \emph{higher} than the Sokolov-Ternov value of $8/(5\sqrt3) \simeq 92.4\%$.

The Derbenev-Kondratenko formula eq.~\eqref{eq:PDK} includes only longitudinal momentum recoils due to photon emissions,
because these are by far the most important in real storage rings.
However, their \emph{formalism} is applicable to treat momentum recoils in all directions, longitudinal and transverse.
\cite{BM} extended the Derbenev-Kondratenko formalism to also include vertical momentum recoils.
They treated the same model as \cite{BL} and obtained eq.~\eqref{eq:PBL},
but with a slightly different expression for $f$, viz.~\cite{BM}, eq.~(40)
\begin{equation}
\label{eq:fBM}
f_{BM} = \frac{2}{\gamma} + \frac{(g-2)Q^2}{Q^2-\nu^2} \,.
\end{equation}
However, the extra term $2/\gamma$ is nonresonant (a constant) and is negligible for $\gamma\gg1$.
The Derbenev-Kondratenko formula eq.~\eqref{eq:PDK}, and the extension to vertical momentum recoils by \cite{BM},
were derived in an inertial reference frame, using lab-frame QED.
Hence the polarization formula obtained by \cite{BL} is equivalent to that obtained using lab-frame QED in an inertial frame.
\cite{JacksonUnruh} published a review of the contributions of \cite{DK73}, \cite{BL} and \cite{BM}.
The contributon of transverse momntum recoils to the polarization is negligible in real storage rings.

For the record, \cite{Manerecoilallplanes} extended the Derbenev-Kondratenko formalism
to include transverse momentum recoils in all planes.
We introduce a triad of vectors $(\bm{d},\bm{e},\bm{f})$ as follows.
First define a set of basis vectors by a right-handed orthonormal triad $(\bm{e}_1,\bm{e}_2,\bm{e}_3)$, 
where $\bm{e}_1$ points radially outwards, $\bm{e}_2$ is longitudinal and counterclockwise and $\bm{e}_3$ points vertically upwards.
The electron velocity vector is $\bm{\beta}=\bm{v}/c=(\beta_1,\beta_2,\beta_3)$.
Then define $\bm{d}=\gamma(\partial{\bm{n}}/d\gamma)$, $\bm{e} = \partial\bm{n}/d\beta_1$ and $\bm{f} = \partial\bm{n}/d\beta_3$.
Recall from eq.~\eqref{eq:PDK} that $\hat{\bm{b}}$ is a unit vector along $\bm{v}\times\dot{\bm{v}}$ and $\rho$ is the local bend radius of the particle orbit.
The asymptotic polarization level is given by
\begin{equation}
  P_{\rm asymp} = \frac{8}{5\sqrt3}\,\frac{P_{\rm num}}{P_{\rm denom}} \,,
\end{equation}
where
\begin{equation}
\begin{aligned}
  P_{\rm num} &= 
  \biggl\langle\frac{1}{|\rho|^3}\biggl(\hat{\bm{b}}\cdot\bm{n} -\hat{\bm{b}}\cdot\bm{d} +\frac{1}{3\gamma}\hat{\bm{v}}\cdot\bm{f}
  \\
  &\qquad\qquad\quad
  -\frac{1}{3\gamma}\bm{n}\cdot(\bm{d}\times\bm{e})\biggr)\biggr\rangle 
\end{aligned}
\end{equation}
and
\begin{equation}
\begin{aligned}
  P_{\rm denom} &= \biggl\langle\frac{1}{|\rho|^3}\biggl(1 -\frac29(\bm{n}\cdot\hat{\bm{v}})^2 +\frac{11}{18}|\bm{d}|^2
  \\
  &\qquad\qquad\quad
  +\frac{13}{90\gamma^2}(|\bm{e}|^2+|\bm{f}|^2)
  \\
  &\qquad\qquad\quad
  +\frac{1}{9\gamma}\frac{\dot{\bm{v}}}{|\dot{\bm{v}}|}\cdot(\bm{n}\times\bm{f})\biggr)\biggr\rangle \,.
\end{aligned}
\end{equation}
As a side note,
\cite{HS} calculated the vacuum electromagnetic fluctuations in a circularly accelerating reference frame.
They discovered the fluctuations have a nonzero Poynting vector in the forward direction (tangent to the orbit).
\cite{ManeHS} showed that the resulting Poynting vector, multiplied by the electromagnetic fine-structure constant $\alpha_{\rm QED}$,
matches the radiated power of classical synchrotron radiation in the forward direction.

\section{Invariant spin field}\label{ch:ISF}
\subsection{General remarks}
In this chapter, we describe algorithms to calculate the $\bm{n}$ axis in real storage rings.
As we have repeatedly noted, $\bm{n}$ is not a single vector, but is a unit vector field over the phase-space manifold of the orbital motion.
The term ``invariant spin field'' (ISF) to denote $\bm{n}$ was introduced by \cite{BarberISF} and the term has since become popular.
It signifies that $\bm{n}$ is a \emph{field}, pertaining to the \emph{spin}, and yields a (dynamical) \emph{invariant}.

Our focus is no longer quantum electrodynamics and photon emission,
but instead the dynamics of particle motion in the prescribed electromagnetic fields of a storage ring.
Also note that, technically, the spin quantization axis $\bm{n}$ is defined for \emph{any} storage ring which circulates particles with spin.
This includes hadron rings such as RHIC, which circulates polarized proton beams.
The determination of $\bm{n}$ (and the off-axis spin tune) is not therefore strictly limited to radiative spin polarization, but is applicable more generally.

To begin, we summarize the orbital motion of the particles in a storage ring.
The reader is referred to many excellent textbooks, or summer school lectures,
for a comprehensive description of the orbital motion of the particles in a storage ring,
e.g.~\cite{EdwardsSyphers}.
For the spin, much of the material below is taken from \cite{MSY2}.
Most storage rings are planar, with a set of magnets to bend (or guide) the particle beam around a closed path.
The reference (or design) orbit typically consists of a set of circular arcs joined by straight lines.
The specification of the reference orbit also includes a reference energy (or momentum).
It is convenient to employ the arc-length $s$ around the ring as the independent variable.
Suppose the ring circumference is $2\pi R$, where $R$ is the mean radius.
Then define a ``generalized azimuth'' $\theta=s/R$, so that $0 \le \theta \le 2\pi$ around the ring.
We shall employ $\theta$ as the independent variable in the equations of motion below
(thereby avoiding confusion between $s$ as the arc-length or the spin).
We employ the following coordinate system, based on the reference orbit.
We denote the basis vectors by a right-handed orthonormal triad $(\bm{e}_1,\bm{e}_2,\bm{e}_3)$, 
where $\bm{e}_1$ points radially outwards, $\bm{e}_2$ is longitudinal and counterclockwise and $\bm{e}_3$ points vertically upwards.
Suppose the local radius of curvature of the reference orbit is $\rho$ (where $1/\rho=0$ in a straight section), then
\begin{equation}
\label{eq:dej/dtheta}
\frac{d\bm{e}_1}{d\theta} = R\frac{\bm{e}_2}{\rho} \,, \quad
\frac{d\bm{e}_2}{d\theta} = -R\frac{\bm{e}_1}{\rho} \,, \quad
\frac{d\bm{e}_3}{d\theta} = 0\,.
\end{equation}
For later use, the radial coordinate variable is denoted by $x$ along $\bm{e}_1$.

Because of inevitable misalignments and tolerances of manufacture of the magnets, the particle orbits are centered around a so-called \emph{closed orbit},
which, as its name suggests, is a closed loop around the ring.
(To put it another way, storage rings are built such that a closed orbit \emph{does} exist.)
The orbital motion then consists of longitudinal and transverse oscillations around the motion of a particle on the closed orbit.
The origin of the phase-space of the orbital oscillations is the closed orbit.
As is standard in the literature, we assume the orbital motion is integrable, and is described by action-angle variables
$(I_j,\phi_j)$, $j=1,2,3$. For brevity we also write $(\bm{I},\bm{\phi})$.
Here $\bm{I}$ denotes the action variables and $\bm{\phi}$ denotes the angle variables.
There are two degrees of freedom for the transverse oscillations and one for the longitudinal oscillations.
For historical reasons, the transverse oscillations are called (horizontal and vertical) \emph{betatron oscillations}
and the longitudinal oscillations are called \emph{synchrotron oscillations}.
By definition, the action variables $\bm{I}$ are dynamical invariants.
For the angle variables, it is conventional to measure the oscillation frequency in units of the orbital revolution frequency.
The resulting dimensionless oscillation frequency is called a \emph{tune},
viz.~the horizontal and vertical betatron tunes and the synchrotron tune.
We denote the orbital tunes by $Q_j$, $j=1,2,3$.
Then $d\phi_j/d\theta = Q_j$, $j=1,2,3$.
In cases where we treat only a single oscillation mode below, we shall drop the subscript $j$.
For later use, we define $\bm{Q}$ as the vector of orbital tunes
and we define $\bm{\mu}=2\pi\bm{Q}$ as the vector of one-turn orbital phase advances.

\subsection{Spin action-angle variables}
The spin action variable is $J = \bm{s}\cdot\bm{n}$, and is conventionally normalized to have the values $\pm1$.
We shall denote the spin tune by $\nu$.
We know that the $\bm{n}$ axis (the invariant spin field) is a function of the orbital phase-space location $(\bm{I},\bm{\phi})$.
What about the spin tune?
From action-angle theory, the tune (whether for the orbit or the spin) must depend on the action variables only.
The spin tune $\nu$ is a function of the orbital action variable $\bm{I}$ only.
\cite{BarberISF} termed it the ``amplitude dependent spin tune'' (ADST) (``amplitude'' referring to the action variables of the orbital motion).

\subsection{Spin precession equation}
We denote the spin precession vector in accelerator coordinates by $\bm{W}$.
Then
\begin{equation}
\label{eq:Wacc}
\begin{split}
\bm{W} &= -R\frac{e}{pc} \biggl(1+\frac{x}{\rho}\biggr)\biggl[(a\gamma+1)\bm{B}
  -\frac{a\gamma^2}{\gamma+1}\bm{\beta}\cdot\bm{B}\,\bm{\beta}
  \\
  &\qquad\qquad\qquad\qquad\quad
  -\biggl(a\gamma+\frac{\gamma}{\gamma+1}\biggr)\bm{\beta}\times\bm{E}\,\biggr]
  \\
  &\qquad
-\frac{\bm{e}_3}{\rho} \,.
\end{split}
\end{equation}
The spin precession equation of motion in accelerator coordinates is
\begin{equation}
\label{eq:BMTacc}
\frac{d\bm{s}}{d\theta} = \bm{W}\times\bm{s} \,.
\end{equation}
In later work, we decompose $\bm{W}=\bm{W}_0+\bm{w}$, where $\bm{W}_0$ refers to spin precession on the closed orbit and $\bm{w}$ is the contribution from the orbital oscillations.

\subsection{Spin basis vectors}
We next define a right-handed orthonormal triad of spin basis vectors $(\bm{l}_0,\bm{m}_0,\bm{n}_0)$ as follows.
Also define $\bm{k}_0 = \bm{l}_0+i\bm{m}_0$.
They satisfy eq.~\eqref{eq:BMTacc} on the closed orbit, i.e.~$d\bm{n}_0/d\theta = \bm{W}_0\times\bm{n}_0$, etc.
Because the spin motion is a rotation, one can say that 
\begin{equation}
\begin{pmatrix}\bm{l}_0 \\ \bm{m}_0 \\ \bm{n}_0 \end{pmatrix}_{\theta+2\pi} = 
M_0(\theta,\theta+2\pi) \begin{pmatrix}\bm{l}_0 \\ \bm{m}_0 \\ \bm{n}_0 \end{pmatrix}_\theta \,. 
\end{equation}
Here $M_0$ is a $3\times 3$ orthogonal matrix which denotes the spin rotation through one turn around the storage ring.
By construction, $M$ has a rotation axis, i.e.~an eigenvector with eigenvalue unity.
That vector is $\bm{n}_0$.
It is the solution for the $\bm{n}$ axis on the closed orbit of a storage ring.
To state it another way, $\bm{n}_0$ is the one-turn periodic solution of the spin precession on the closed orbit:
$\bm{n}(\theta+2\pi) = \bm{n}(\theta)$.
Hence to solve for $\bm{n}$ on the closed orbit, we compute the one-turn spin rotation matrix and find its rotation axis.
As for the other spin basis vectors, they have the property
$\bm{k}(\theta+2\pi) = e^{-2\pi\nu_0}\bm{k}(\theta)$.
Here $\nu_0$ is the spin tune on the closed orbit.
The vectors $\bm{l}_0$ and $\bm{m}_0$ execute a counterclockwise rotation through an angle $2\pi\nu_0$ around $\bm{n}_0$, for one circuit around the sing.
It is straightforward to solve eq.~\eqref{eq:BMTacc} on the closed orbit of a model of a planar ring with vertical magnetic fields on the closed orbit
(joined by straight sections, with zero magnetic field).
Then $\bm{n}_0=\bm{e}_3$ is vertical, as expected, and the spin tune on the closed orbit is $\nu_0=a\gamma_0$.

\subsection{Energy calibration}
\emph{Why is the relation $\nu_0=a\gamma_0$ in a planar ring so important?}
It is an experimental fact that frequencies, in general, can be measured very accurately.
Since $\gamma_0 = E_0/(mc^2)$, one can calibrate the beam energy of a storage ring by measuring the value of $\nu_0$ and computing $E_0 = \nu_0(mc^2/a)$.
The particle mass $m$ and the value of $a$ are known to high precision.
(For electrons, $mc^2/a \simeq 440.65$ MeV.)
Measuring the spin tune is most accurate known technique to calibrate the beam energy of a storage ring.
As stated in the Introduction, the mass of the $Z^0$ boson ($91.1876$ GeV) was measured to an accuracy of $\pm2.1$ MeV \cite{PDG_Z0}.
The measurement of the mass of the $Z^0$ boson was one of the highlights of the LEP physics program \cite{Zedometry}.
See also \cite{LEPenergy} for a review of the precise calibration of the LEP beam energy.

\subsection{Off-axis motion: periodicities}
To determine the $n$ axis away from the closed orbit, we require the following.
First, $\bm{n}$ must satisfy eq.~\eqref{eq:BMTacc}.
We have repeatedly noted that the $\bm{n}$ axis is a function of the orbital phase-space location $(\bm{I},\bm{\phi})$.
Since $(I_j,\phi_j)$ and $(I_j,\phi_j+2\pi)$ represent the same phase-space point,
and also because $\theta$ and $\theta+2\pi$ represent the same location around the ring,
i.e.~the same phase space,
then $\bm{n}$ must moreover satisfy the following periodicity relations:
\begin{equation}
\label{eq:nper}
\begin{split}
  &\bm{n}(I_1,I_2,I_3,\phi_1,\phi_2,\phi_3,\theta)
  \\
  &\qquad= \bm{n}(I_1,I_2,I_3,\phi_1,\phi_2,\phi_3,\theta+2\pi)
  \\
  &\qquad= \bm{n}(I_1,I_2,I_3,\phi_1+2\pi,\phi_2,\phi_3,\theta)
  \\
  &\qquad= \bm{n}(I_1,I_2,I_3,\phi_1,\phi_2+2\pi,\phi_3,\theta)
  \\
  &\qquad= \bm{n}(I_1,I_2,I_3,\phi_1,\phi_2,\phi_3+2\pi,\theta) \,.
\end{split}
\end{equation}
On the closed orbit, where $I_1=I_2=I_3=0$, eq.~\eqref{eq:nper} reduces to simply $\bm{n}_0(\theta) = \bm{n}_0(\theta+2\pi)$, as we have already seen.
Up to a minus sign, the relations in eq.~\eqref{eq:nper} determine $\bm{n}$ uniquely.

\subsection{Single Resonance Model}
\begin{figure}[!htb]
\centering
\includegraphics[width=0.45\textwidth]{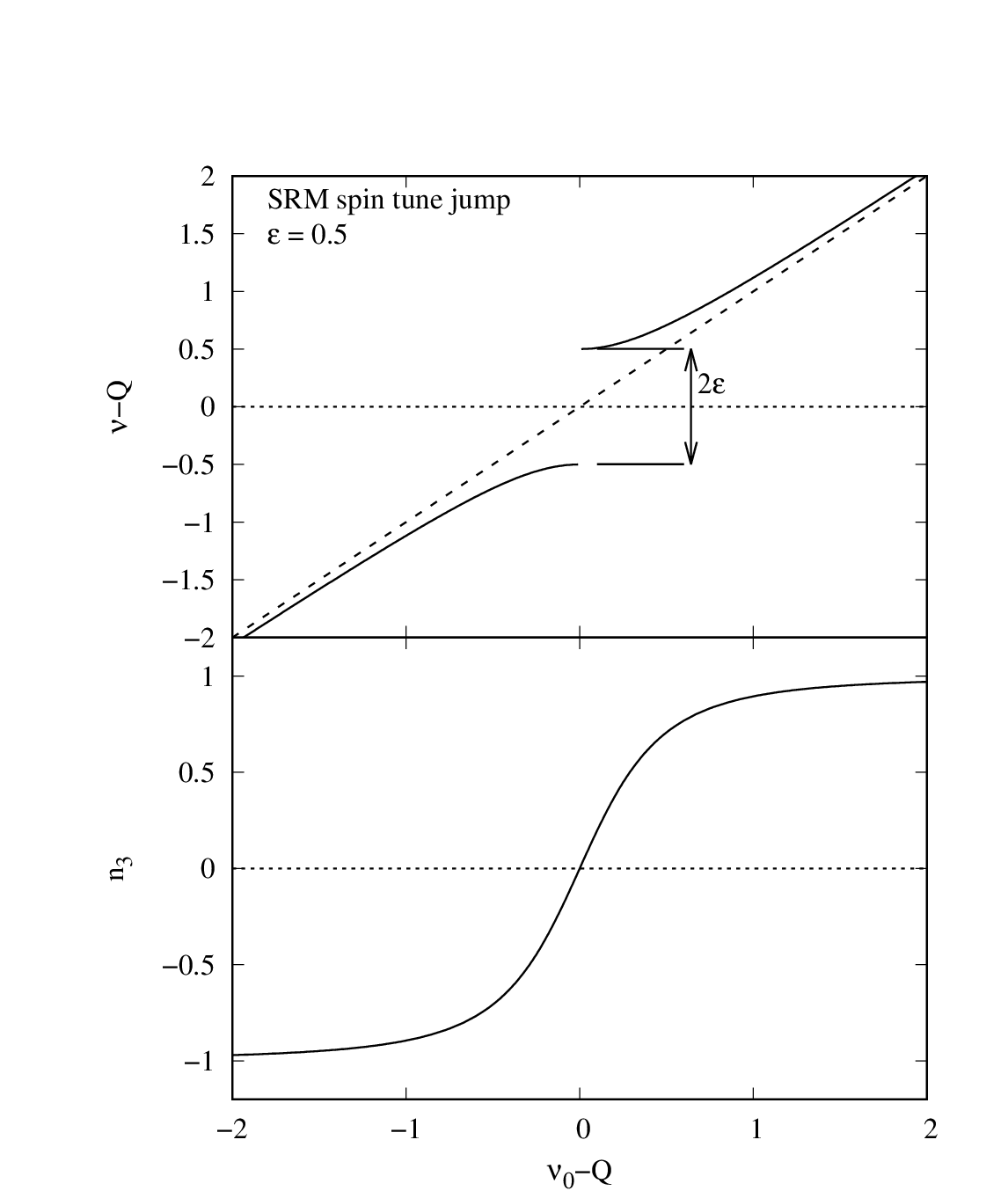}
\caption[\;Graph of spin tune and $\bm{n}$ axis for the Single Resonance Model]{
\label{fig:SRM_graph}
Top panel: graph of $\nu-Q$ vs.~$\nu_0-Q$, where $\nu$ is the spin tune, $\nu_0$ is the unperturbed spin tune
and $Q$ is the tune of the orbital mode.
The arrow indicates the jump in the value of the spin tune at the spin resonance $\nu_0=Q$.
The dashed curve plots the value of $\nu_0-Q$.
Bottom panel: graph of $n_3$, the vertical component of the spin quantization axis $\bm{n}$, vs.~$\nu_0-Q$.}
\end{figure}

The expression for $\bm{n}$ can be solved exactly in the case where $\bm{w}$ contains only a single term.
This is called the Single Resonance Model (SRM).
It is a good approximation when only one off-axis term dominates the spin precession.
The value of $\bm{W}$ is from a uniform vertical magnetic field and the perturbing term $\bm{w}$ lies in the horizontal plane.
The orbital motion has a tune $Q$, with corresponding angle variable $\phi$ (so $d\phi/d\theta=Q$).
The amplitude of the off-axis term in $\bm{w}$ is $\varepsilon$, a real-valued constant.
The spin precession vector is
\begin{equation}
\label{eq:Wsrm}  
  \bm{W}_{\rm SRM} = \nu_0\bm{e}_3 +\varepsilon\,(\bm{e}_1\cos\phi +\bm{e}_2\sin\phi) \,.
\end{equation}
The solution for $\bm{n}$ is
\begin{equation}
\label{eq:nsrm}  
  \bm{n}_{\rm SRM} = \frac{(\nu_0-Q)\bm{e}_3 +\varepsilon\,(\bm{e}_1\cos\phi +\bm{e}_2\sin\phi)}{\sqrt{(\nu_0-Q)^2+\varepsilon^2}} \,.
\end{equation}
The corresponding value of the spin tune is, with the convention $\nu_{\rm SRM}=\nu_0=Q$ if $\nu_0=Q$,
\begin{equation}
\label{eq:nusrm}  
  \nu_{\rm SRM} = Q +\frac{\nu_0-Q}{|\nu_0-Q|}\sqrt{(\nu_0-Q)^2+\varepsilon^2} \,.
\end{equation}
Observe that, for fixed $\varepsilon \ne0$, the value of $\bm{n}_{\rm SRM}$ is a continuous function of $\nu_0-Q$.
However, the spin tune $\nu_{\rm SRM}$ is \emph{not}.
A graph of the spin tune $\nu$ (actually $\nu-Q$) and the vertical component $n_3$ is plotted as a function of $\nu_0-Q$ Fig.~\ref{fig:SRM_graph}.
The resonance driving term is $\varepsilon=0.5$.
The spin tune exhibits a ``jump'' (discontinuity) of magnitude $2\varepsilon$ as the spin resonance at $\nu_0=Q$ is crossed.
Observe that, for fixed $\varepsilon\ne0$, if we sweep the value of $\nu_0-Q$ from a large negative value to a large positive value,
then $\bm{n}_{\rm SRM}$ changes (or ``flips'') orientation from vertically down ($\nu_0-Q<0$ and $|\nu_0-Q| \gg \varepsilon$)
to vertically up ($\nu_0-Q>0$ and $|\nu_0-Q| \gg \varepsilon$).
The relevant theory was worked out by \cite{FS}.
This fact has been employed in so-called ``spin flippers'' to reverse the direction of a stored polarized beam \emph{in situ}.
The perturbing term in $\bm{w}$ is generated by a so-called ``spin flipper'' and dominates the value of $\bm{w}$.
The value of $\nu_0-Q$ is swept across the resonance, by varying the value of $Q$ in the spin flipper.
Spin flipping was first performed at the Budker Institute of Nuclear Physics (BINP) \cite{PoluninShatunov}.
Spin flippers have since been employed with success in many storage rings, including hadron rings. See \cite{MSY2} for references.

\subsection{Perturbative algorithms}
\subsubsection{Basic equation}
We now solve for $\bm{n}$ away from the closed orbit, for more general values of $\bm{w}$.
We begin with perturbative algorithms.
Recall that we decompose $\bm{W}=\bm{W}_0+\bm{w}$, where $\bm{W}_0$ refers to spin precession on the closed orbit and $\bm{w}$ is the contribution from the orbital oscillations.
We have already introduced the spin basis vectors $(\bm{l}_0,\bm{m}_0,\bm{n}_0)$ and $\bm{k}_0=\bm{l}_0+i\bm{m}_0$.
We follow the formalism by \cite{YokoyaSyncSdb}.
We introduce a complex variable $\zeta$ and parameterize $\bm{n}$ as follows.
\begin{equation}
\label{eq:nzeta}
\bm{n} = \sqrt{1-|\zeta|^2}\,\bm{n}_0 + \Re(\zeta\bm{k}_0^*) \,.
\end{equation}
Obviously $\zeta=0$ on the closed orbit. The equation of motion for $\zeta$ is
\begin{equation}
\label{eq:eomzeta}
\frac{d\zeta}{d\theta} = -i\bm{w}\cdot\bm{k}_0\sqrt{1-|\zeta|^2}+i\bm{w}\cdot\bm{n}_0\zeta \,.
\end{equation}
This is a nonlinear differential and is too complicated to solve, in general.

\subsubsection{SLIM}
We begin by neglecting $\zeta$ on the right-hand side of eq.~\eqref{eq:eomzeta}.
This yields the SLIM algorithm by \cite{ChaoSLIM}.
It is a linear response algorithm, because Chao also linearized the orbital equations of motion around the closed orbit.
We obtain
\begin{equation}
\frac{d\zeta}{d\theta} \simeq -i\bm{w}\cdot\bm{k}_0\,.
\end{equation}
The formal solution which satisfies the periodicities in eq.~\eqref{eq:nper} is 
\begin{equation}
\label{eq:zetasol}
\zeta = -i\int_{-\infty}^\theta \bm{w}\cdot\bm{k}_0\,d\theta^\prime \,.
\end{equation}
It is explained by \cite{ChaoSLIM} (see also \cite{MSY2}) how to render this into a well-defined expression.
Let us simplify the problem and treat only one orbital oscillation mode, with tune $Q$, write
$u = -i\bm{w}\cdot\bm{k}_0$, where $u$ has the one-turn periodicity $u(\theta+2\pi) = e^{i2\pi(\nu_0+Q)}$.
Then we employ an infinitesimal damping factor to obtain a convergent sum as follows:
\begin{equation}
\label{eq:zetamode}
\begin{split}
  \zeta &= \int_{-\infty}^\theta u\,d\theta^\prime
  \\
  &= \biggl(\int_\theta^{\theta+2\pi} u\,d\theta^\prime\biggr) \Bigl(e^{-i2\pi(\nu_0+Q)}
  \\
  &\qquad\qquad\qquad\qquad\quad
  +e^{-i4\pi(\nu_0+Q)} +\cdots\Bigr)
  \\
  &= \frac{1}{e^{i2\pi(\nu_0+Q)}-1} \int_\theta^{\theta+2\pi} u\,d\theta^\prime \,.
\end{split}
\end{equation}
Hence it is only necessary to compute a one-turn integral around the ring, from $\theta$ to $\theta+2\pi$.
Observe that the expression for $\zeta$ diverges if the denominator vanishes, i.e.~if $\nu_0+Q = r$, where $r$ is an integer.
This is called a ``resonance denominator'' in the literature.
It yields an example of a spin resonance: the $n$ axis is ill-defined (and the polarization level drops to zero).
The SLIM algorithm yields expressions for the strengths of first order spin resonances,
i.e.~$\nu_0\pm Q_j = r$, $j=1,2,3$.

\subsubsection{Spin resonance strength}
Note that eq.~\eqref{eq:zetamode} yields an off-axis expression for $\bm{n}$.
The actual spin resonance strength is given by the phase-space derivative $\gamma(\partial\bm{n}/d\gamma)$
(see eq.~\eqref{eq:PDK}).
We eschew the technical details to calculate $\gamma(\partial\bm{n}/d\gamma)$ here;
they can be found in the original literature, e.g.~\cite{YokoyaSyncSdb}, or the review by \cite{MSY2}.
From the definition of the off-axis spin precession vector $\bm{w}$ and eq.~\eqref{eq:Wacc},
we see that, in first-order perturbation theory,
$\gamma(\partial\bm{n}/d\gamma)$ is proportional to $(a\gamma+1)$ times the gradient of the magnetic field $\bm{B}$.
Symbolically,
\begin{equation}
  \gamma\frac{\partial\bm{n}}{d\gamma} \propto (a\gamma+1)\times \Bigl(\gamma\frac{\partial\bm{B}}{d\gamma}\Bigr) +\dots 
\end{equation}
This led to the scaling formula by \cite{Buonscaling} in eq.~\eqref{eq:Pscaling} in Chapter \ref{ch:DK}.

\subsubsection{Sideband spin resonances}
Our treatment follows \cite{YokoyaSyncSdb}, which furnishes a more complete analysis in an earlier paper by \cite{DKS}.
Let us simplify eq.~\eqref{eq:eomzeta} as follows. We neglect only the term in $|\zeta|^2$ to obtain 
\begin{equation}
\label{eq:eomzetasdb}
\frac{d\zeta}{d\theta} \simeq -i\bm{w}\cdot\bm{k}_0+i\bm{w}\cdot\bm{n}_0\zeta \,.
\end{equation}
Eq.~\eqref{eq:eomzetasdb} is also a linear differential equation, hence can be solved using standard techniques.
The formal solution which satisfies the periodicities in eq.~\eqref{eq:nper} is 
\begin{equation}
\label{eq:zetasdbformal}
\zeta = -i\,e^{-i\chi(\theta)}\int_{-\infty}^\theta e^{i\chi(\theta^\prime)}\bm{w}\cdot\bm{k}_0\,d\theta^\prime \,.
\end{equation}
Here $\chi(\theta)$ is given by
\begin{equation}
\chi(\theta) = -\int_{-\infty}^\theta \bm{w}\cdot\bm{n}_0\,d\theta^\prime \,.
\end{equation}
The full details are given by \cite{YokoyaSyncSdb}.
Here, we simplify the exposition by assuming $\bm{w}\cdot\bm{k}_0$ contains only a single orbital mode, with tune $Q_1$,
and $\bm{w}\cdot\bm{n}_0$ also contains only a single orbital mode, with tune $Q_2$.
Let us approximate
\begin{equation}
\chi(\theta) \simeq \int_{-\infty}^\theta A\cos\phi_2 \,d\theta^\prime = \frac{A}{Q_2}\,\sin\phi_2 \,.
\end{equation}
Here $A$ is a phenomenological constant.
Then $e^{i\chi(\theta)} = e^{i(A/Q_2)\sin\phi_2}$.
We employ the Jacobi-Anger identity to express this as a weighted sum of Bessel functions
\begin{equation}
e^{i\chi(\theta)} = e^{i(A/Q_2)\sin\phi_2} = \sum_{m=-\infty}^\infty e^{im\phi_2} J_m(A/Q_2) \,.
\end{equation}
Then, omitting unnecessary nonresonant factors which merely clutter the notation (here $\phi_2^\prime = \phi_2(\theta^\prime)$),
\begin{equation}
\label{eq:zetasdb}
\begin{split}
  \zeta(\theta) &\propto -i\int_{-\infty}^\theta \sum_{m=-\infty}^\infty e^{im\phi_2^\prime} J_m(A/Q_2) \bm{w}\cdot\bm{k}_0 \,d\theta^\prime
  \\
  &\propto -i\sum_{m=-\infty}^\infty J_m(A/Q_2) \int_{-\infty}^\theta e^{im\phi_2^\prime} \bm{w}\cdot\bm{k}_0 \,d\theta^\prime
  \\
  &\propto -i\sum_{m=-\infty}^\infty \frac{J_m(A/Q_2)}{e^{i2\pi(\nu_0+Q_1+mQ_2)}-1} \;\times
  \\
  &\qquad\qquad\qquad
  \int_\theta^{\theta+2\pi} e^{im\phi_2^\prime} \bm{w}\cdot\bm{k}_0\,d\theta^\prime \,.  
\end{split}
\end{equation}
Compared to eq.~\eqref{eq:zetamode}, the expression for $\zeta$ in eq.~\eqref{eq:zetasdb} contains an infinite sum of spin resonance denominators.
The expression for $\zeta$ diverges when $\nu_0+Q_1+mQ_2=r$, where $m$ and $r$ are integers.
The spin resonances are called ``sideband'' resonances of the ``parent'' spin resonance (which is indexed by $m=0$).
The strengths of the sideband spin resonances are proportional to Bessel functions.

\subsubsection{Synchrotron sideband resonances}
The expressions for $\gamma(\partial\bm{n}/d\gamma)$ and the spin resonance strengths,
for both the parent and sideband resonances, are given by \cite{YokoyaSyncSdb}.
As a practical matter, the most important sideband spin resonances are driven by the synchrotron oscillations.
The strengths of the synchrotron sideband resonances are characterized by the following parameter \cite{YokoyaSyncSdb}, eq.~(3.8).
\begin{equation}
\sigma = \frac{1}{Q_s}\Bigl(\gamma\frac{\partial \nu}{\partial\gamma}\Bigr)_0\,\frac{\sigma_E}{E} \,.
\end{equation}
Here $Q_s$ is the synchrotron tune (the tune of the synchrotron oscillations), $\sigma_E$ is the rms beam energy spread and $E$ is the reference beam energy.
Also $\gamma(d\nu/d\gamma)$ is the phase-space derivative of the spin tune $\nu$,
analogous to $\gamma(d\bm{n}/d\gamma)$ for the $\bm{n}$ axis.
The subscript ``$0$'' indicates the expression is evaluated on the closed orbit.
In the case of a planar ring, $\nu = a\gamma$ and $(\gamma(d\nu/d\gamma))_0 = a\gamma_0$. Then we obtain
\begin{equation}
\label{eq:tunemodindex}  
\sigma = \frac{a\gamma_0}{Q_s}\,\frac{\sigma_E}{E} \,.
\end{equation}
The parameter $\sigma^2$ is called the \emph{tune modulation index}.
The synchrotron sideband resonance strengths were given by
\cite{YokoyaSyncSdb}, eq.~(3.17) and \cite{DKS}, eq.~(5.2).
\begin{equation}
\label{eq:syncsdb}
\begin{split}
  \biggl\langle\biggl|\gamma\frac{\partial{n}}{d\gamma}\biggr|\biggr\rangle &=
  |a_n|^2 \sum_{m=-\infty}^\infty \biggl[\frac{\Delta\nu}{(\Delta\nu +mQ_s)^2 - Q_s^2}\biggr]^2\;\times
  \\
  &\qquad\qquad\qquad\qquad
  e^{-\sigma^2}I_m(\sigma^2) \,.
\end{split}
\end{equation}
Here $a_n$ is the amplitude of the parent resonance, centered at an integer $n$.
Also $\Delta\nu = \nu_0-n$ is assumed to have a small magnitude.
Next, $I_m$ is a modified Bessel function of the first kind.

The synchrotron sideband resonances grow in strength as the beam energy increases.
In modern storage rings, $\sigma_E/E \propto E$, hence in a planar ring, for fixed $Q_s$, $\sigma \propto E^2$ or $\sigma^2\propto E^4$.
It was the increasing strength of the synchrotron sideband resonances, as the beam energy increased,
which ultimately invalidated the scaling formula by \cite{Buonscaling} in eq.~\eqref{eq:Pscaling}.
See \cite{ManeFCC} for a summary of the details of such a calculation.

\subsubsection{SMILE}
The SMILE algorithm by \cite{ManeSMILE} treats all the higher order spin resonances.
The orbital equations of motion are also linearized around the closed orbit.
We parameterize $\bm{n}$ via $\bm{n} = n_1\bm{l}_0+n_2\bm{m}_0+n_3\bm{n}_0$.
Also define $w_1 = \bm{w}\cdot\bm{l}_0$, $w_2 = \bm{w}\cdot\bm{m}_0$ and $w_3 = \bm{w}\cdot\bm{n}_0$.
Then
\begin{equation}
\begin{split}
  \frac{d\ }{d\theta}\begin{pmatrix}n_1 \\ n_2 \\ n_3\end{pmatrix}
    &= \begin{pmatrix} 0 & -w_3 & w_2 \\ w_3 & 0 & -w_1 \\ -w_2 & w_1 & 0 \end{pmatrix} \begin{pmatrix}n_1 \\ n_2 \\ n_3\end{pmatrix}
      \\
&= -i(\bm{w}\cdot\bm{J}) \begin{pmatrix}n_1 \\ n_2 \\ n_3\end{pmatrix} \,.
\end{split}
\end{equation}
Here $\bm{J}$ is a vector of three spin 1 angular momentum matrices.
The formal solution which satisfies the periodicities in eq.~\eqref{eq:nper} is 
\begin{equation}
  \begin{pmatrix}n_1 \\ n_2 \\ n_3\end{pmatrix}_\theta
    = \textrm{T}\biggl\{\exp\biggl(-i\int_{-\infty}^\theta (\bm{w}\cdot\bm{J})\,d\theta^\prime\biggr)\biggr\} \begin{pmatrix}0 \\ 0 \\ 1\end{pmatrix} \,.
\end{equation}
Here ``$\textrm{T}$'' denotes a $\theta$-ordered product of operators.
As with eq.~\eqref{eq:zetasol}, an infinitesimal damping factor is required to render the integral well-defined.
There are resonance denominators whenever $\nu_0 +r_1Q_1 +r_2Q_2+r_3Q_3=r$, where $r$, $r_1$, $r_2$ and $r_3$ are integers,
i.e.~spin resonances of all orders.
The SMILE algorithm was historically the first algorithm to calculate the $n$ axis systematically to all orders of perturbation theory,
yielding expressions for the strengths of higher order spin resonances.
\cite{ManePAC89} employed the SMILE algorithm to successfully fit the electron polarization measurements at SPEAR by the SLAC-Wisconsin collaboration \cite{SPEARpol}.
See Fig.~\ref{fig:SPEAR_fit}.

\subsubsection{Lie algebra}
\cite{YokoyaLieAlg} employed the formalism of Lie algebra to publish an algorithm to calculate the $n$ axis to all orders of perturbation theory
(in powers of the orbital action variables)
\emph{without} linearizing the orbital equations of motion around the closed orbit.
It was historically the first algorithm to calculate the $n$ axis to all orders of perturbation theory
without linearizing the orbital equations of motion around the closed orbit.
However, the formalism is mathematically sophisticated and too complicated to explain here.
The fundamental idea is to construct a generating function (perturbatively) to perform successive canonical transformations to
express the Hamiltonian in joint spin-orbit action-angle form.
The spin action variable yields the $\bm{n}$ axis.
Yokoya's algorithm also yields the amplitude dependent spin tune.
The SLIM and SMILE algorithms compute only $\bm{n}$ but not the amplitude dependent spin tune.

\subsection{Nonperturbative algorithms}
\subsubsection{Stroboscopic averaging}
Stroboscopic averaging \cite{StrobAvg} is an elegant numerical technique to calculate $\bm{n}$.
The key idea is this: note that $\bm{s}\cdot\bm{n}$ is invariant for motion on any orbital trajectory,
and all other values for the spin $\bm{s}$ precess around $\bm{n}$.
Hence, in the long term, only the component $\bm{s}\cdot\bm{n}$ of any spin vector $\bm{s}$ on that trajectory will survive;
the components of $\bm{s}$ orthogonal to $\bm{n}$ will average to zero.
This, of course, presumes that the orbit is not on a spin resonance.
We therefore determine $\bm{n}$ in the following way. 
We first select a base azimuth $\theta_*$.
We parameterize the orbital motion using action-angle variables $(\bm{I},\bm{\phi}_*)$.
We hold the actions $\bm{I}$ at fixed values and do not mention them explicitly below.
We then launch a particle at the phase-space point $(\bm{\phi}_* - j\bm{\mu}, \theta_* - 2j\pi)$,
with the spin $\bm{s}_j$ oriented along $\bm{n}_0$ and track the spin-orbit motion for $j$ turns.
By construction, the particle will end up at the phase-space point $(\phi_*,\theta_*)$ for any $j=1,2,\dots$.
We perform this procedure for $N$ spins, setting $j = 1, 2,\dots,N$, i.e. one particle is tracked
for one turn, another is tracked for two turns, etc.
It is possible with some caching of information to make the computational complexity $O(N)$ not $O(N^2)$.
At the end, we obtain a set of $N$ spins $\bm{s}_j$, $j = 1, 2,\dots,N$, all at the phase-space point $(\phi_*,\theta_*)$.
We now \emph{average} the spins to obtain
\begin{equation}
\label{eq:strobavg}
\bar{\bm{s}} = \frac{1}{N}\sum_{j=1}^N \bm{s}_j \,.
\end{equation}
As already explained, the key idea is that the (unknown) components orthogonal to $\bm{n}$ average to zero for sufficiently large $N$.
However, the (also unknown) components $\bm{s}\cdot\bm{n}$ are invariant during the tracking and will sum up to a nonzero average value.
Hence, for sufficiently large $N$ and if the
average converges then $\bar{\bm{s}}\parallel\bm{n}$.
Hence we set $\bm{n} = \bar{\bm{s}}/|\bar{\bm{s}}|$.
The term ``stroboscopic'' arises from the Poincar{\'e} sections in the tracking:
we observe the spins only at discrete intervals, not continuously.

Note that the sum in eq.~\eqref{eq:strobavg} might not converge if $\bm{n}_0\perp \bm{n}$, or the rate of convergence may be poor, if the orbit is close to a spin resonance.
The choice of launching the spins pointing along $\bm{n}_0$ is merely a simple convention and a good choice if the orbit is far from a spin resonance resonance.
However, another initial vector for the $\bm{s}_j$ could be employed.

\subsubsection{SODOM2}
The SODOM2 algorithm was published by \cite{YokoyaSODOM2}.
It superseded an earlier algorithm SODOM by \cite{YokoyaSODOM}, which we do not discuss here.
In SODOM2, the spin is represented by a two-component spinor $\psi$.
For brevity, we omit explicit mention of the action variables $\bm{I}$ and write only the angle variables $\bm{\phi}$.
Also for brevity we define $\bm{\mu}=2\pi\bm{Q}$.
As with stroboscopic averaging, we select a reference phase-space point $(\bm{\phi}_*,\theta_*)$.
We also omit explicit mention of the base azimuth $\theta_*$ below.
The one-turn map equation for a spinor $\psi$ representing $\bm{n}$ is
\begin{equation}
\label{eq:SODOM2v}
M\psi(\bm{\phi}_*) = e^{-iv(\bm{\phi}_*)/2}\psi(\bm{\phi}_*+\bm{\mu}) \,.
\end{equation}
Here $v$ is a periodic function of $\bm{\phi}_*$, i.e.~$v(\bm{\phi}_* + 2\pi) = v(\bm{\phi}_*)$.
We find another function $u(\bm{\phi}_*)$, also $2\pi$ periodic in $\bm{\phi}_*$, such that
\begin{equation}
v(\bm{\phi}_*) + u(\bm{\phi}_*+\bm{\mu}) - u(\bm{\phi}_*) = \mu_s \,.
\end{equation}
Here $\mu_s$ is independent of $\bm{\phi}_*$ and $\theta_*$, and depends only on the actions $\bm{I}$.
Then we define a new spinor $\Psi = e^{iu(\bm{\phi}_*)/2}\psi$.
Then
\begin{equation}
\label{eq:SODOM2u}
M\Psi(\bm{\phi}_*) = e^{-i\mu_s/2}\Psi(\bm{\phi}_*+\bm{\mu}) \,.
\end{equation}
We now expand $M$ and $\Psi$ as Fourier series in $\bm{\phi}_*$, viz.
\begin{subequations}
\begin{align}
  M &= \sum_{\bm{m}} M_{\bm{m}}e^{i\bm{m}\cdot\bm{\phi}_*} \,,
  \\
\Psi &= \sum_{\bm{m}} \Psi_{\bm{m}}e^{i\bm{m}\cdot\bm{\phi}_*} \,.
\end{align}
\end{subequations}
Then eq.~\eqref{eq:SODOM2u} can be reexpressed as 
\begin{equation}
\label{eq:SODOMeq}
e^{-i\bm{m}\cdot\bm{\mu}}\sum_{\bm{m}^\prime}M_{\bm{m}-\bm{m}^\prime}\Psi_{\bm{m}^\prime} = e^{-i\mu_s/2}\Psi_{\bm{m}} \,.
\end{equation}
This can be visualized as an infinite-dimensional matrix eigenvalue problem for $e^{-i\mu_s/2}$.
The eigenvector elements are the Fourier harmonics $\Psi_{\bm{m}}$.
The solution for $\bm{n}$ is $\bm{n} = \Psi^\dagger\bm{\sigma}\Psi$, where $\bm{\sigma}$ is a vector of Pauli matrices.
The solution for $\bm{n}$ is the same for all the eigenvector solutions of eq.~\eqref{eq:SODOMeq}.
The amplitude dependent spin tune is given by $\nu = \mu_s/(2\pi)$.
The solutions for the spin tune are not unique up to $\mu_s \to \mu_s + \bm{m}\cdot\bm{\mu}$, or $\nu \to \nu +\bm{m}\cdot\bm{Q}$.
However, this is a known ambiguity in the formal canonical transformation theory and not a flaw in the SODOM2 formalism.
In practice, the matrix $M$ must be truncated to a finite number of Fourier harmonics, which
generates numerical error in the solution, and the eigenvalue calculation must also be performed numerically.
The bulk of the computation time is spent, not surprisingly, in the eigenvalue/eigenvector solver.

\subsubsection{MILES}
The name MILES \cite{ManeMILES} is simply an anagram of the name of the earlier formalism
SMILE by \cite{ManeSMILE}. There is, otherwise, no connection between the two formalisms. The
fact that $\bm{n}$ is a unit vector field over the orbital phase-space means that its transformation is given by
$\bm{\sigma}\cdot\bm{n}(z_f) = M \bm{\sigma}\cdot\bm{n}(z_i)M^{-1}$,
where $z$ symbolically denotes a point in the orbital phase-space and $\bm{\sigma}$ is a vector of Pauli matrices.
Hence
\begin{equation}
\label{eq:MILESsign}
\bm{\sigma}\cdot{n} = \begin{pmatrix} n_3 & n_1-in_2 \\ n_1+in_2 & -n_3 \end{pmatrix} \,.
\end{equation}
We shall employ the definitions $n_\pm = n_1 \pm in_2$ below.
We omit explicit mention of the orbital action variables $\bm{I}$ below.
Let the orbital angles at the base azimuth $\theta_*$ be $\bm{\phi}_*$.
Denote the one-turn orbital phase advances by $\bm{\mu}=2\pi\bm{Q}$.
The one-turn mapping for $\bm{n}$, at the azimuth $\theta_*$, is then
\begin{equation}
\label{eq:MILESmap}
\bm{\sigma}\cdot\bm{n}(\bm{\phi}_*+\bm{\mu}) = M \bm{\sigma}\cdot\bm{n}(\bm{\phi}_*)M^{-1} \,.
\end{equation}
Eq.~\eqref{eq:MILESmap} provides a complete, nonperturbative technique to calculate $\bm{n}$.
An arbitrary SU(2) matrix $M$ can be parametrized as follows:
\begin{equation}
\label{eq:MILES_MSU2}
M = \begin{pmatrix} f & -g^* \\ g & f^* \end{pmatrix} \,.
\end{equation}
Here $f$ and $f$ are functions of the orbital motion and $ff^*+gg^*=1$.
In terms of $f$ and $g$, we deduce
\begin{equation}
\label{eq:MILESeq}
\begin{split}
  n_3(\bm{\phi}_*+\bm{\mu}) &= (ff^*-gg^*)n_3(\bm{\phi}_*) 
  \\
  &\qquad
  -f^*g^*n_+(\bm{\phi}_*) -fgn_-(\bm{\phi}_*) \,,
  \\
  n_+(\bm{\phi}_*+\bm{\mu}) &= 2f^*gn_3(\bm{\phi}_*)
  \\
  &\qquad
  +f^{*2}n_+(\bm{\phi}_*) -g^2n_-(\bm{\phi}_*) \,.
\end{split}
\end{equation}
We now expand $f$ and $g$, and also $n_3$ and $n_\pm$, as Fourier series in $\bm{\phi}_*$ and equate terms.
The procedure is similar to the SODOM2 algorithm described above.
As with the SODOM2 algorithm, the resulting equations constitute an infinite set of relations, in general.
Hence for practical computation they must be truncated to a finite set.

\subsubsection{Fiber bundle}
More abstract later developments treat the spin as a fiber bundle over the phase-space manifold of particle orbits \cite{FiberBundle}.
The mathematics is unfortunately too advanced to be described here.
The authors treated both spin $\frac12$ and spin $1$ particles.
For spin $1$, the polarization density matrix has both rank 1 and rank 2 spherical harmonics
(or vector and tensor polarization components).
All the other formalisms above treated only spin $\frac12$ and vector polarization.

\subsection{Spin matching/spin transparency}
We remarked earlier that for a model of motion in a uniform vertical magnetic field, the $\bm{n}$ axis is vertical at all points in the orbital phase-space,
hence $\gamma(\partial\bm{n}/\partial\gamma)$ equals zero.
Hence there are no depolarizing spin resonances in such a model.
The storage ring HERA was a proton-lepton (electron or positron) collider at DESY in Hamburg, Germany.
The lepton ring was configured to deliver \emph{longitudinal} polarization at the interaction points (the H1, ZEUS and HERMES detectors).
This was accomplished by the installation of so-called ``spin rotators'' to rotate the lepton polariztion from vertical, in the ring arcs,
to longitudinal at the interaction points.
However, if the ring design is modified so that $\bm{n}$ is no longer parallel to the magnetic guide field in the ring arcs,
the set of $\bm{n}$ axes develops a (large) spread in the orbital phase-space.
This causes the magnitude of $\gamma(\partial\bm{n}/\partial\gamma)$ to become large,
resulting in a serious reduction of the polarizaton level (see the Derbenev-Kondratenko formula eq.~\eqref{eq:PDK}).
Special steps must be taken to configure the ring design to minimize the depolarization caused by $\gamma(\partial\bm{n}/\partial\gamma)$.
The procedure is called \emph{spin matching}.
It is also called \emph{spin transparency}.
Early papers to introduce this concept were \cite{ChaoYokoya1981} and \cite{Yokoya1981} (where it was termed ``spin transparency'').
In the case of HERA, \cite{BuonSteffen} designed a so-called ``mini'' spin rotator and steps to counteract the reduction in the asymptotic polarization level.
\cite{BarberHERApol} reported the successful results obtained at HERA.
The details of spin matching are too technical to describe here,
but spin matching remains an important procedure for storage rings which circulate polarized particle beams.

The foregoing discussion of spin matching is called ``strong spin matching'' because it requires a modification to the basic design/layout of the storage ring.
An alternative technique called ``harmonic spin matching'' has also been employed with success.
The basic idea is to Fourier analyze the driving terms of the (first-order) spin resonances and to
employ orbit corrector magnets to eliminate the Fourier harmonic which makes the strongest contribution to the depolarization.
In practice, one compensates both the real and imaginary parts (or sine and cosine terms) in that Fourier harmonic.
Harmonic spin matching is an empirical procedure, to adjust a set of corrector magnets, without modifying the basic ring layout.
It was employed with success at PETRA, increasing the measured polarization from approximately $25\%$ to over $80\%$ \cite{RossmanithHSM}, Fig.~8.
Harmonic spin matching was later employed with success at LEP,
increasing the asymptotic polarization level from approximately $10\%$ to over $40\%$.
See Fig.~5 in \cite{LEPHSM}, 
which displays the 
asymptotic polarization level attained at LEP during its operational history up to 1993.
Note the reference to ``asymptotic'' because the actual measured polarization was lower, and the results were extrapolated to $t\to\infty$
(see eq.~\eqref{eq:Pt}).

More recently, \cite{SH_BAGELS2} have introduced a program ``BAGELS'' (``Best Adjustment Groups for ELectron Spin'')
with new ideas for spin matching.
This is a ``coordinate-space'' technique as oppose to Fourier analysis.
The idea is to apply local closed orbit bumps to minimize the depolarization.
The technique is still a theoretical proposal at the time of the writing of this article, but theoretical simulations hold promise.

\subsection{Topological quantum phase}
The spin rotations induced by spin rotators cause a shift of the spin tune.
This is because the spins traverse a path on a ``unitary sphere'' which includes a topological quantum phase in addition to the dynamical phase advance.
Hence the spins traverse an extra angle \emph{(which could be negative)} in one circuit around a storage ring,
and this is manifested experimentally as a shift of the spin tune.
This is called ``Berry's phase'' \cite{Berry} or more generally an ``Aharanov-Anandan phase''\cite{AA}.
See \cite{ManeAA} for the application to spin motion in storage rings.

\section*{Acknowledgments}
There are several people I would like to thank, as mentors and friends who have helped and educated me in this subject over the years.
I am deeply indebted to all of them.
However none of them are responsible for the contents of this article; all the limitations in the text are solely mine.
In alphabetical order, they are
Desmond Barber,
John Bell,
Yaroslav Derbenev,
Alex Dragt,
Michael Fisher,
Bernard Gittelman,
Robert Glasser,
Kurt Gottfried,
Louis Hand,
John David Jackson,
Anatoliy Kondratenko,
Ivan Koop,
Waldo MacKay,
James Niederer,
Alexey Otboev,
Rajendran Raja,
Hikaru Sato,
Charles Sinclair,
Klaus Steffen,
Yuri Orlov,
Yuri Shatunov,
Jorg Wenninger
and
Kaoru Yokoya.


\end{document}